# DNS Study on Physics of Late Boundary Layer Transition

Chaoqun Liu[1] and Ping Lu[2]

University of Texas at Arlington, Arlington, Texas 76019, USA
cliu@uta.edu

**This paper serves as a review of our recent new DNS study on physics of late boundary layer transition. This includes mechanism of the large coherent vortex structure formation, small length scale generation and flow randomization. The widely spread concept “vortex breakdown to turbulence”, which was considered as the last stage of flow transition, is not observed and is found theoretically incorrect. The classical theory on boundary layer transition is challenged and we proposed a new theory with five steps, i.e. receptivity, linear instability, large vortex formation, small length scale generation, loss of symmetry and randomization to turbulence. We have also proposed a new theory about turbulence generation. The new theory shows that all small length scales (turbulence) are generated by shear layer instability which is produced by large vortex structure with multiple level vortex rings, multiple level sweeps and ejections, and multiple level negative and positive spikes near the laminar sub-layers. Therefore, “turbulence” is not generated by “vortex breakdown” but rather positive and negative spikes and consequent high shear layers. “Shear layer instability” is considered as the “mother of turbulence”. This new theory may give a universal mechanism for turbulence generation and sustenance – the energy is brought by large vortex structure through multiple level sweeps not by “vortex breakdown”. The ring-like vortex with or without legs is found the only form existing inside the flow field. The first ring-like vortex shape, which is perfectly circular and perpendicularly standing, is the result of the interaction between two pairs of counter-rotating primary and secondary streamwise vortices. Multiple rings are consequences of shear layer instability produced by momentum deficit which is formed by vortex ejections. The U-shaped vortices are not new but existing coherent large vortex structure. Actually, the U-shaped vortex, which is not a secondary but tertiary vortex, serves as an additional channel to support the multiple ring structure. The loss of symmetry and randomization are caused by internal property of the boundary layer. The loss of symmetry starts from the second level ring cycle in the middle of the flow field and spreads to the bottom of the boundary layer and then the whole field. More other new physics have also been discussed. Finally, classical theory such as “Richardson energy cascade” and “Kolmogorov small length scale” is revisited and challenged.**

## Nomenclature

$M_\infty$ = Mach number

$Re$ = Reynolds number

$_{in}$ = inflow displacement thickness

$T_w$ = wall temperature

---

[1] Professor, AIAA Associate Fellow, University of Texas at Arlington, USA, AIAA Associate Fellow
[2] PhD Student, AIAA Student Member, University of Texas at Arlington, USA

$T_\infty$ = free stream temperature

$Lz_{in}$ = height at inflow boundary

$Lz_{out}$ = height at outflow boundary

$Lx$ = length of computational domain along x direction

$Ly$ = length of computational domain along y direction

$x_{in}$ = distance between leading edge of flat plate and upstream boundary of computational domain

$A_{2d}$ = amplitude of 2D inlet disturbance

$A_{3d}$ = amplitude of 3D inlet disturbance

= frequency of inlet disturbance

$_{2d}$, $_{3d}$ = two and three dimensional streamwise wave number of inlet disturbance

= spanwise wave number of inlet disturbance

$R$ = ideal gas constant

= ratio of specific heats

$_\infty$ = viscosity

# I. Introduction

Tubulence is still covered by a mystical veil in nature after over a century of intensive study. Following comments are made by wekipedia web page at http://en.wikipedia.org/wiki/Turbulence "Nobel Laureate Richard Feynman described turbulence as "the most important unsolved problem of classical physics" (USA Today 2006). According to an apocryphal story, Werner Heisenberg (another Nobel Prize Winner) was asked what he would ask God, given the opportunity. His reply was: "When I meet God, I am going to ask him two questions: Why relativity? And why turbulence? I really believe he will have an answer for the first." (Marshak, 2005). Horace Lamb was quoted as saying in a speech to the British Association for the Advancement of Science, "I am an old man now, and when I die and go to heaven there are two matters on which I hope for enlightenment. One is quantum electrodynamics, and the other is the turbulent motion of fluids. And about the former I am rather optimistic" (Mullin 1989; Davidson 2004).

These comments clearly show that the mechanism of turbulence formation and sustenance is still a mystery for research. Note that both Heisenberg and Lamb were not optimistic for the turbulence study.

## 1.1 Richardson's vortex and energy cascade theory (1928)

Richardson believed that a turbulent flow is composed by "eddies" of different sizes. The large eddies will be stretching, unstable and breaking up to smaller eddies. These smaller eddies undergo the same process, giving rise to even smaller eddies. This process will continue until reaching a sufficiently small length scale such that the viscosity of the fluid can effectively dissipate the kinetic energy into internal energy. During the process of vortex breakdown, the kinetic energy of the initial large eddy is divided into the smaller eddies.

## 1.2 Kolmogorov assumption (1941)

The classical theory on turbulence was given by Kolmogorov, a famous Russian mathematician. In general, the large scales of a flow are not isotropic, because they are determined by the particular boundary conditions. Agreeing with Richardson, Kolmogorov assumed: in the Richardson's energy cascade, this geometrical and directional information is lost, while the scale is reduced and so that the statistics of the small scales has a universal character: they are statistically isotropic for all turbulent flows when the Reynolds number is sufficiently high. It was assumed that there is no dissipation during the energy transfer from large vortex to small vortex through "vortex breakdown".

## 1.3 Kolmogovor's first and second hypotheses (1941)

Based on his assumption, Kolmogorov (1941) further gave very famous theories on smallest length scale, which is later called Kolmogorov scale (first hypothesis), and turbulence energy spectrum (second hypothesis):

$$=(\frac{\;^{3}}{\;})^{\frac{1}{4}},$$

$$E(k)=C\;\;^{2/3}k^{-5/3}\text{ and}$$

$$=\;\{2\overline{(\frac{\partial u_1}{\partial x_1})^2}+2\overline{(\frac{\partial u_2}{\partial x_2})^2}+2\overline{(\frac{\partial u_3}{\partial x_3})^2}+\overline{(\frac{\partial u_2}{\partial x_1}+\frac{\partial u_1}{\partial x_2})^2}+\overline{(\frac{\partial u_3}{\partial x_2}+\frac{\partial u_2}{\partial x_3})^2}+\overline{(\frac{\partial u_1}{\partial x_3}+\frac{\partial u_3}{\partial x_1})^2}\}$$

Where, η is Kolmogorov scale, ν is kinematic viscosity, ε is the rate of turbulence dissipation, E is the energy spectrum function, C is a constant and κ is the wave number. These formulas were obtained by Kolmogorov's hypotheses that the small length scales are determined by ν and ε, and E is related to κ and ε. These formulas are unique according to the dimensional analysis (Frisch, 1995).

### 1.4 A short review of study on late boundary layer transition

The transition process from laminar to turbulent flow in boundary layers is a basic scientific problem in modern fluid mechanics. After over a hundred of years of study on flow transition, the linear and weakly non-linear stages of flow transition are pretty well understood (Herbert.1988; Kachanov. 1994). However, for late non-linear transition stages, there are still many questions remaining for research (Kleiser et al, 1991; Sandham et al, 1992; U. Rist et al 1995, Borodulin et al, 2002; Bake et al 2002; Kachanov, 2003). Adrian (2007) described hairpin vortex organization in wall turbulence, but did not discuss the sweep and ejection events and the role of the shear layer instability. Wu and Moin (2009) reported a new DNS for flow transition on a flat plate. They did obtain fully developed turbulent flow with structure of forest of ring-like vortices by flow transition at zero pressure gradients. However, they did not give the mechanism of the late flow transition. The important mechanism of boundary layer transition such as sweeps, ejections, positive spikes, etc. cannot be found from that paper. Recently, Guo et al (2010) conducted an experimental study for late boundary layer transition in more details. They concluded that the U-shaped vortex is a barrel-shaped head wave, secondary vortex, and is induced by second sweeps and positive spikes.

In order to get deep understanding the mechanism of the late flow transition in a boundary layer and physics of turbulence, we recently conducted a high order direct numerical simulation **(**DNS) with 1920×241×128 gird points and about 600,000 time steps to study the mechanism of the late stages of flow transition in a boundary layer at a free stream Mach number 0.5 (Chen et al., 2009, 2010a, 2010b, 2011a, 2011b; Liu et al., 2010a, 2010b, 2010c, 2011a, 2011b, 2011c, 2011d; Lu et al., 2011a, 2011b, 2011c). The work was supported by AFOSR, UTA, TACC and NSF Teragrid. A number of new observations are made and new mechanisms are revealed in late boundary layer transition (LBLT) including:

- **Mechanism on secondary and tertiary vortex formation**
- **Mechanism on first ring-like vortex formation**
- **Mechanism on second sweep formation**
- **Mechanism on high share layer formation**
- **Mechanism on positive spike formation**
- **Mechanism on multiple ring formation**
- **Mechanism on U-shaped vortex formation**
- **Mechanism on small length vortices generation**
- **Mechanism on multiple level high shear layer formation**
- **Mechanism of energy transfer paths**
- **Mechanism on symmetry loss or so called "flow randomization"**
- **Mechanism on thickening of turbulence boundary layer**
- **Mechanism of high surface friction of turbulent flow**

A $\Lambda_2$ technology developed by Jeong and Hussain (1995) is used for visualization.

### 1.5 Different physics observed by our high order DNS

According to our recent DNS, "vortex breakdown" is not observed (Liu et al, 2010a, 2011c, 2011d). Liu gave a new theory that "turbulence is not generated by vortex breakdown but shear layer instability" (Liu et al, 2010a, 2011d; Lu et al, 2011b). As we believe, "**shear layer instability**" is the "**mother of turbulence**"

As Richardson's energy cascade and Kolmogorov's assumption about "vortex breakdown" are challenged, Kolmogorov's assumption that the smallest vortices are statistically isotropic becomes questionable. Since the smallest vortices are generated by the "shear layer instability," which is closely related to the shape of body configuration, they cannot be isotropic.

If all small vortices are generated by "shear layer instability" but not "vortex breakdown", Kolmogorov's first hypothesis will lose the background. The smallest length scales will be related to stability of the weakest unstable "shear layer". This will require a deep study of "shear layer instability" not only "dimensional analysis" (we will publish this analysis soon).

### 1.6 Questions to classical theory on boundary layer transition

The classical theory, which considers "vortex breakdown" as the last stage of boundary layer transition on a flat plate, is challenged and the phenomenon of "hairpin vortex breakdown to smaller structures" is not observed by our new DNS (Liu et all, 2010a, 2011a, 2011b). The so-called "spikes" are actually a process of multi-bridge or multi-ring formation, which is a rather stable large vortex structure and can travel for a long distance.

As experiment is quite expensive and has very limited power in date acquisition, direct numerical simulation (DNS) becomes a more and more important tool to discover physics. The purpose of this work is to find physics of turbulence by direct numerical simulation. The paper is organized as follows: Section I is a background review; Section II shows the case set up and code validation; Section III is our observation and analysis; Section IV is a summary of our new finding; Section V is the conclusions which are made based on our recent DNS.

### 1.7 New theory on boundary layer transition by C. Liu

Classical theory on boundary layer transition can be described by four stages: 1) Boundary layer receptivity; 2) Linear instability; 3) Non-linear growth; 4) Vortex breakdown to turbulence. Apparently, we disagree with the classical theory on "vortex breakdown to turbulence". The new theory of boundary layer transitioncan be described by five stages: 1) Boundary layer receptivity; 2) Linear instability; 3) Large vortex structure formation; 4) Small vortices generation; 5) Symmetry loss and turbulence formation. By the way, the vortex cascade in turbulence given by Richardson, Kolmogorov and others is not observed.

### 1.8 Summary of the new theory on turbulence generation by C. Liu

The new theory on turbulence formation and sustenance shows that all small length scales (turbulence) are generated by shear layer instability which is produced by large vortex structure with multiple level vortex rings, multiple level sweeps and ejections, and multiple level negative and positive spikes near the laminar sub-layers. Therefore, "turbulence" is not generated by "vortex breakdown" but rather positive and negative spikes and consequent high shear layers. "Shear layer instability" is considered as the "mother of turbulence". This new theory may give a universal mechanism for turbulence generation and sustenance – the energy is brought by large vortex structure through multiple level sweeps

Of course, the new theory has to be further studied and confirmed. More mathematical and numerical study is needed

## II. Case Setup and Code Validation

### 2.1 Case setup

The computational domain is displayed in Figure 1. The grid level is 1920×128×241, representing the number of grids in streamwise ($x$), spanwise ($y$), and wall normal ($z$) directions. The grid is stretched in the normal direction and uniform in the streamwise and spanwise directions. The length of the first grid interval in the normal direction at the entrance is found to be 0.43 in wall units ($Y^+$=0.43). The parallel computation is accomplished through the Message Passing Interface (MPI) together with domain decomposition in the streamwise direction (Figure 2). The flow parameters, including Mach number, Reynolds number, etc are listed in Table 1. Here, $x_{in}$ represents the distance between leading edge and inlet, $Lx$ , $Ly$ , $Lz_{in}$ are the lengths of the computational domain in x-, y-, and z-directions, respectively, and $T_w$ is the wall temperature.

Table 1: Flow parameters

| $M_\infty$ | $Re$ | $x_{in}$ | $Lx$ | $Ly$ | $Lz_{in}$ | $T_w$ | $T_\infty$ |
|---|---|---|---|---|---|---|---|
| 0.5 | 1000 | 300.79 $_{in}$ | 798.03 $_{in}$ | 22 $_{in}$ | 40 $_{in}$ | 273.15K | 273.15K |

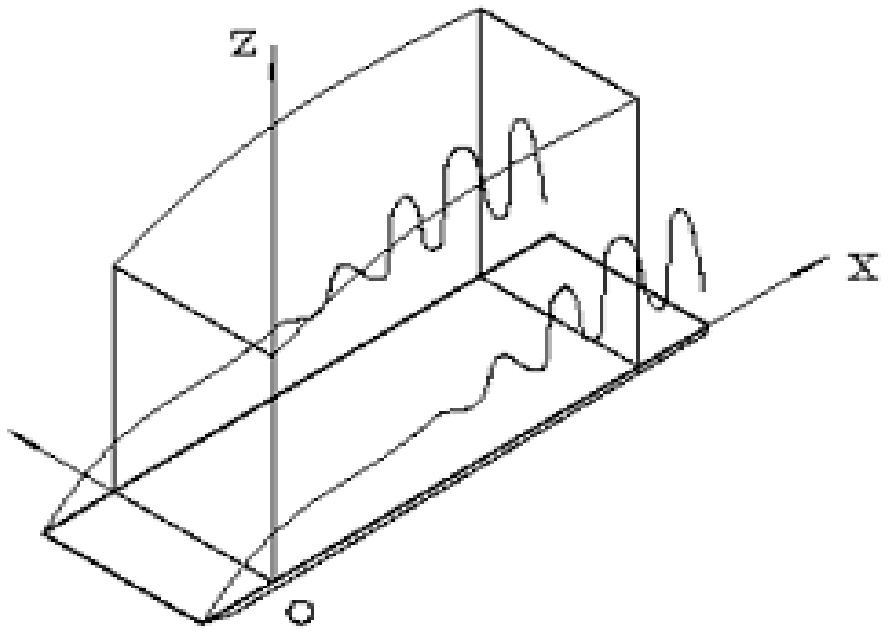

Figure 1: Computation domain

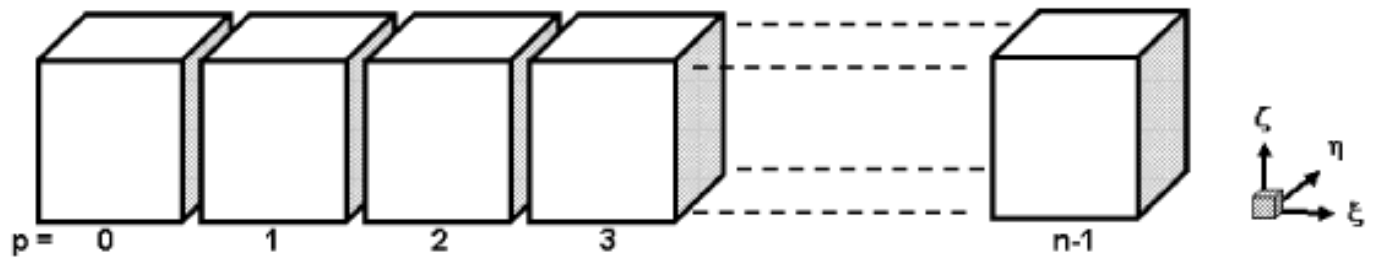

Figure 2: Domain decomposition along the streamwise direction in the computational space

### 2.2 Code Validation

The DNS code – "DNSUTA" has been validated by NASA Langley and UTA researchers (Jiang et al, 2003; Liu et al, 2010a; Lu et al 2011b) carefully to make sure the DNS results are correct.

### 2.2.1 Comparison with Linear Theory

Figure 3 compares the velocity profile of the T-S wave given by our DNS results to linear theory. Figure 4 is a comparison of the perturbation amplification rate between DNS and LST. The agreement between linear theory and our numerical results is quite good.

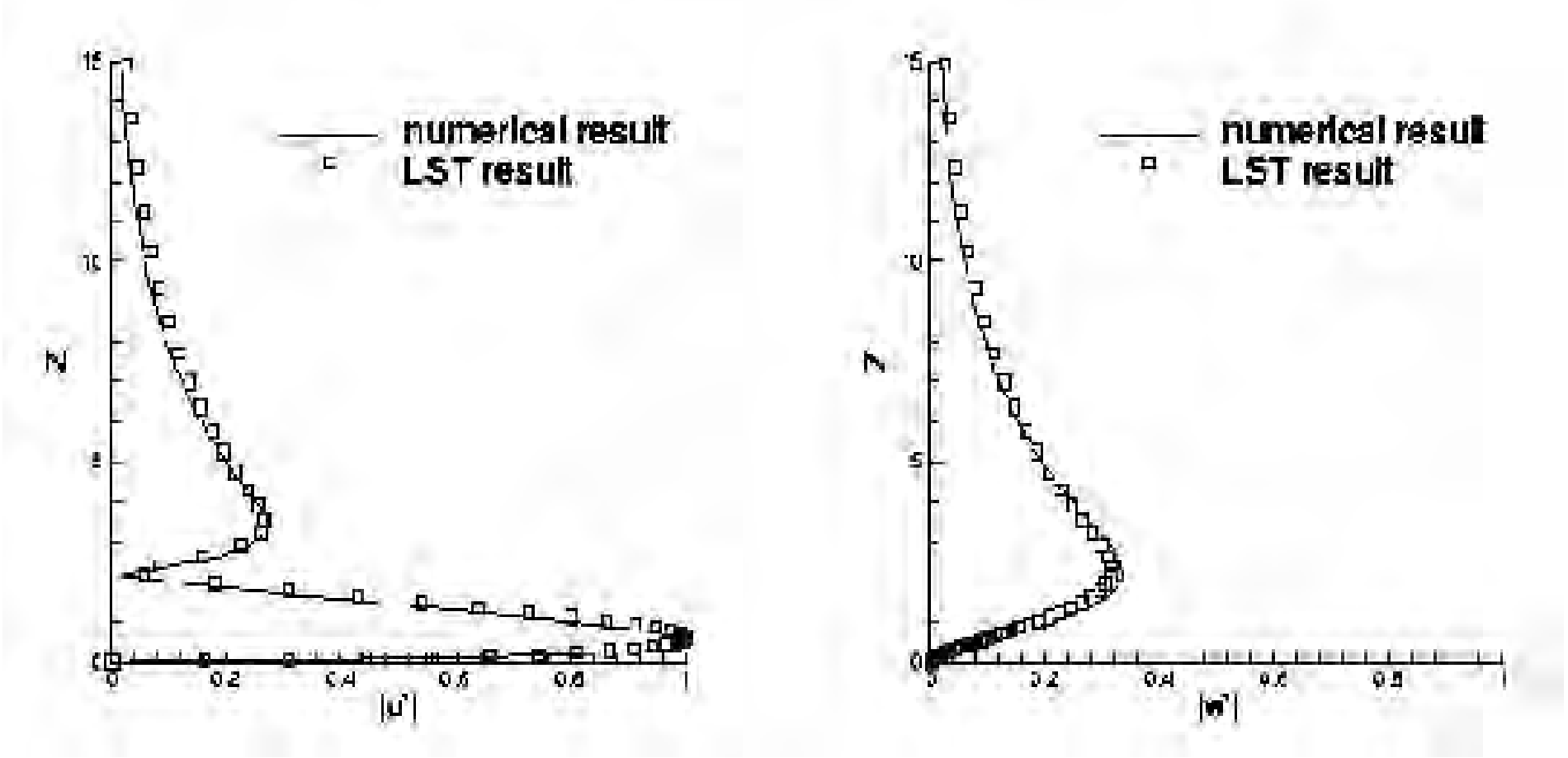

Figure 3: Comparison of the numerical and LST velocity profiles at Rex=394300

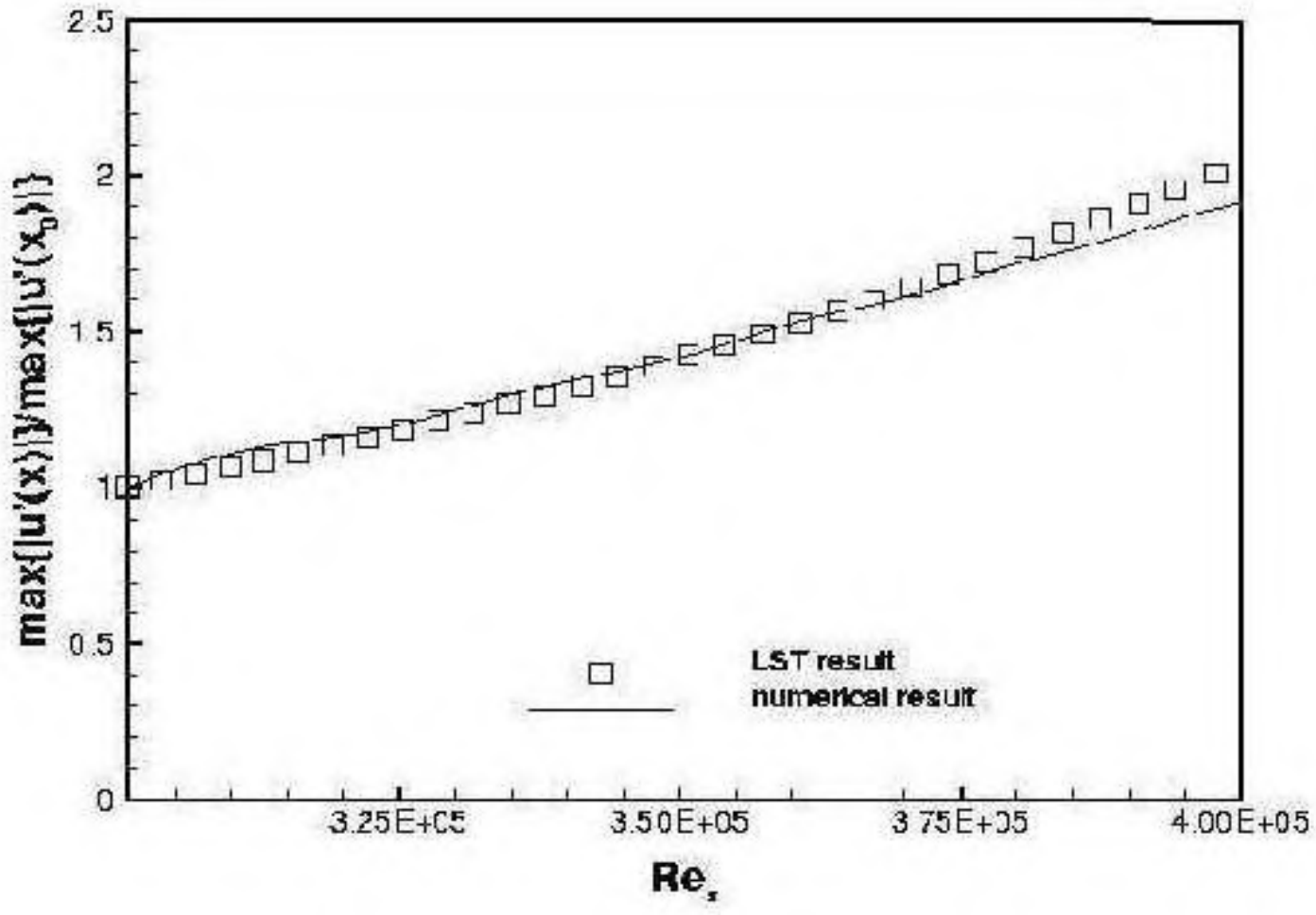

Figure 4: Comparison of the perturbation amplification rate between DNS and LST

**2.2.2 Grid Convergence**

The skin friction coefficient calculated from the time-averaged and spanwise-averaged profile on a coarse and fine grid is displayed in Figure 5. The spatial evolution of skin friction coefficients of laminar flow is also plotted out for comparison. It is observed from these figures that the sharp growth of the skin-friction coefficient occurs after $x \approx 450_{in}$, which is defined as the 'onset point'. The skin friction coefficient after transition is in good agreement with the flat-plate theory of turbulent boundary layer by Cousteix in 1989 (Ducros, 1996). Figures 5(a) and 5(b) also show that we get grid convergence in skin friction coefficients.

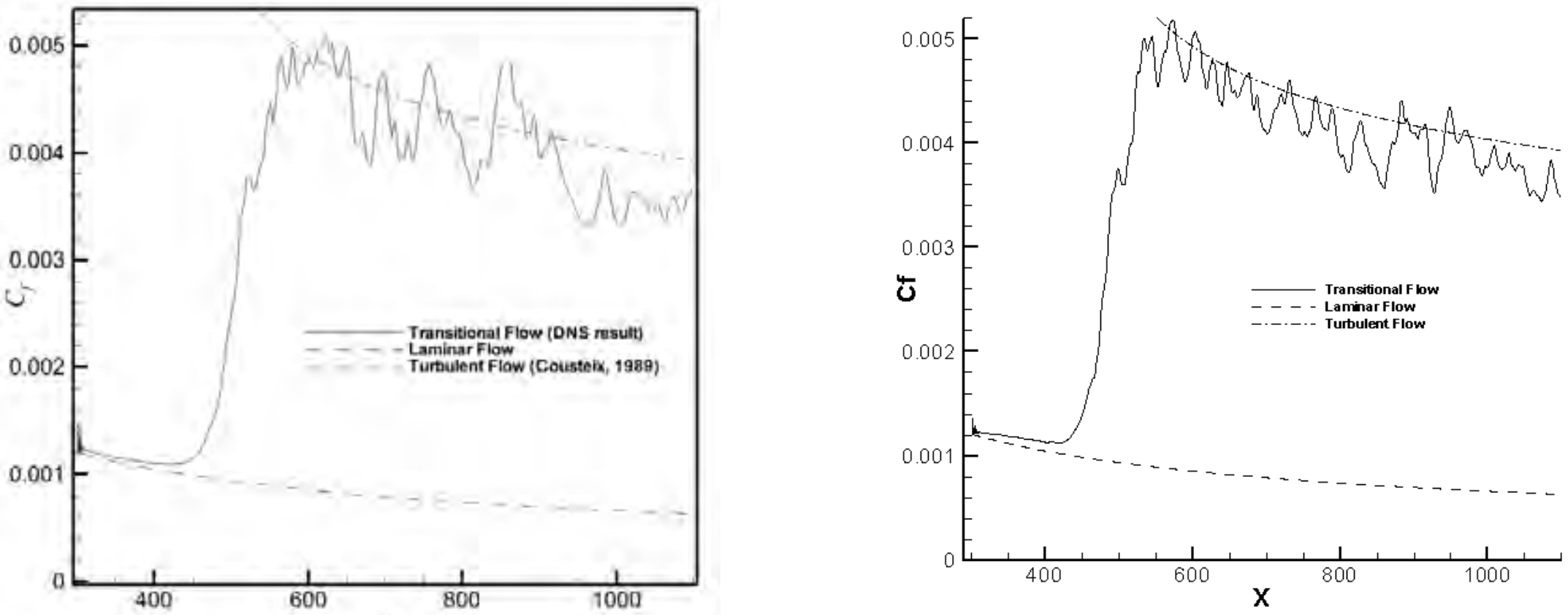

(a) Coarse Grids ( $960 \times 64 \times 121$ ) (b)Fine Grids (1920x128x241)

Figure 5: Streamwise evolutions of the time-and spanwise-averaged skin-friction coefficient

### 2.2.3 Comparison with Log Law

Time-averaged and spanwise-averaged streamwise velocity profiles for various streamwise locations in two different grid levels are shown in Figure 6. The inflow velocity profiles at $x = 300.79 \ _{in}$ is a typical laminar flow velocity profile. At $x = 632.33 \ _{in}$, the mean velocity profile approaches a turbulent flow velocity profile (Log law). This comparison shows that the velocity profile from the DNS results is turbulent flow velocity profile and the grid convergence has been realized.

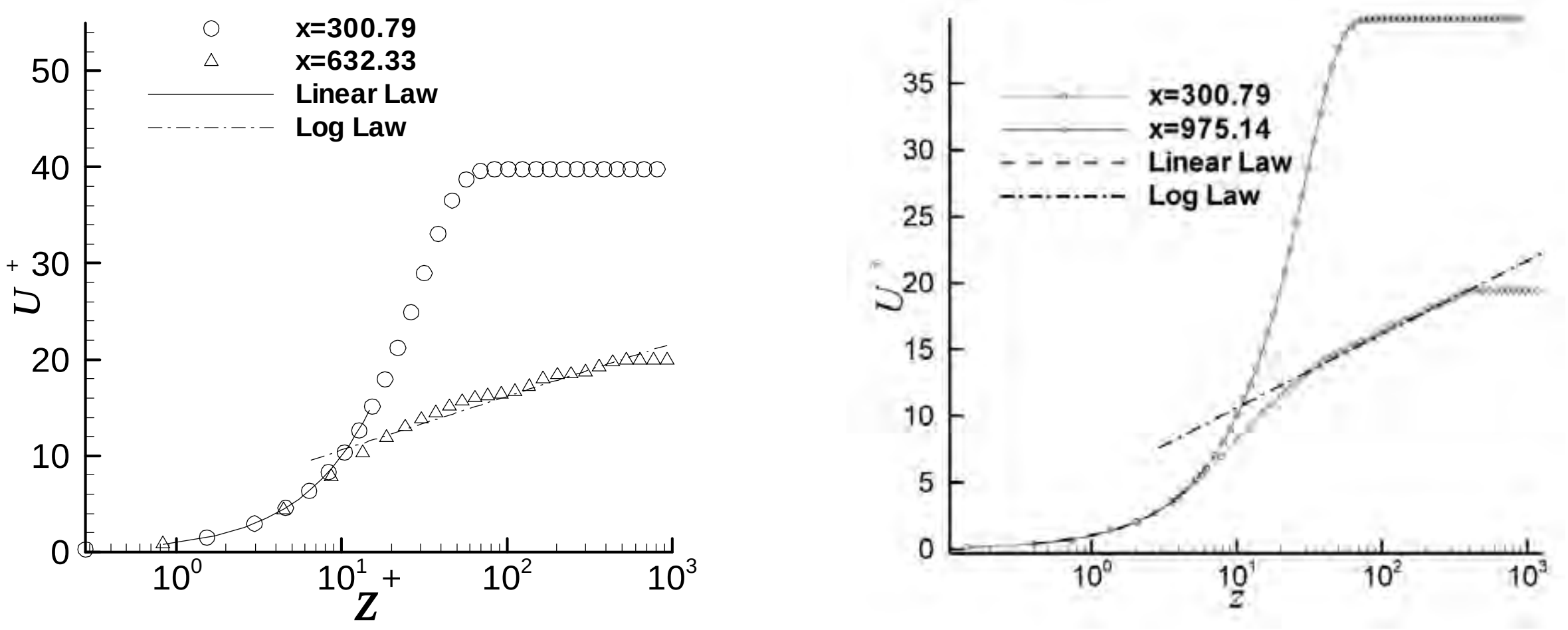

(a) Coarse Grids (960x64x121) (b) Fine Grids (1920x128x241)

Figure 6: Log-linear plots of the time-and spanwise-averaged velocity profile in wall unit

### 2.2.4 Spectra and Reynolds stress (velocity) statistics

Figure 7 shows the spectra in x- and y- directions. The spectra are normalized by z at location of $\mathrm{Re}_x = 1.07 \times 10^6$ and $y^+ = 100.250$. In general, the turbulent region is approximately defined by $y^+ > 100$ and $y/ \ < 0.15$. In our case, The location of $y/ \ = 0.15$ for $\mathrm{Re}_x = 1.07 \times 10^6$ is corresponding to $y^+ \approx 350$ , so the points at $y^+ = 100$ and $250$ should be in the turbulent region. A straight line with slope of -3/5 is also shown for comparison. The spectra tend to tangent to the $k^{-\frac{5}{3}}$ law.

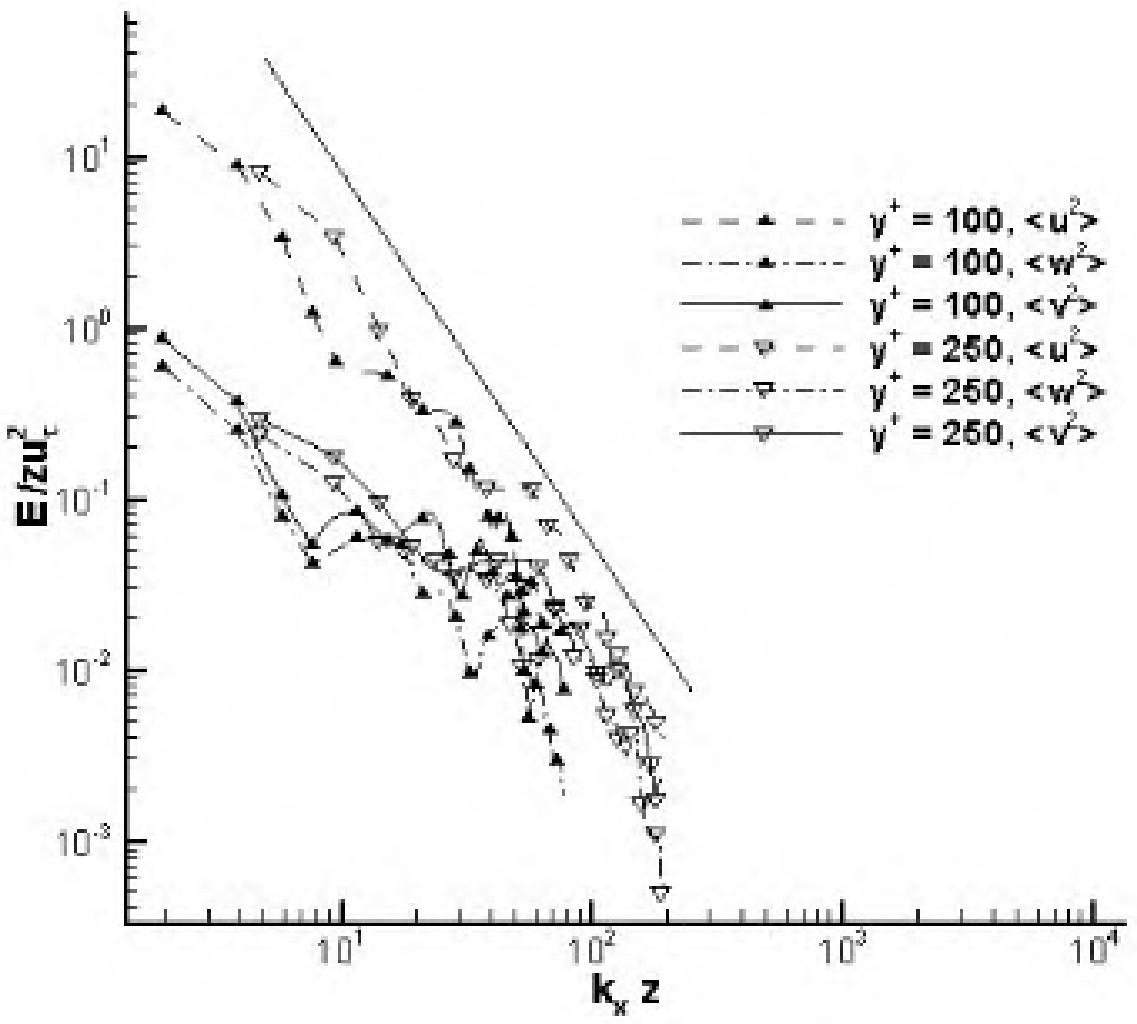

Figure 7(a): Spectra in x- direction

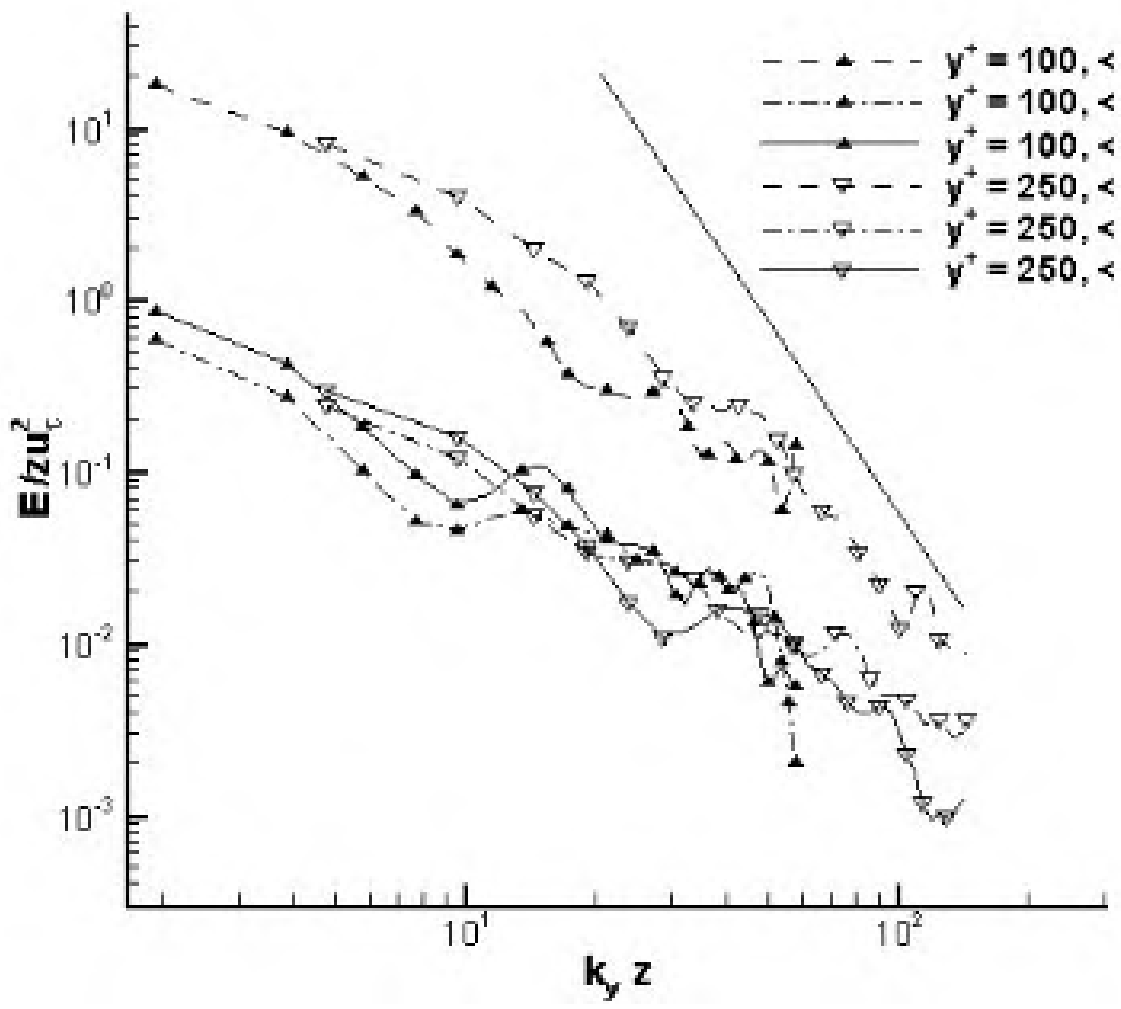

Figure 7(b): Spectra in y- direction

Figure 8 shows Reynolds shear stress profiles at various streamwise locations, normalized by square of wall shear velocity. There are 10 streamwise locations starting from leading edge to trailing edge are selected. As expected, close to the inlet at $\mathrm{Re}_x = 326.8\times10^3$ where the flow is laminar, the values of the Reynolds stress is much smaller than those in the turbulent region. The peak value increases with the increase of $x$. At around $\mathrm{Re}_x = 432.9\times10^3$, a big jump is observed, which indicates the flow is in transition. After $\mathrm{Re}_x = 485.9\times10^3$, the Reynolds stress profile becomes close to each other in the turbulent region. So for this case, we can consider that the flow transition starts after $\mathrm{Re}_x = 490\times10^3$.

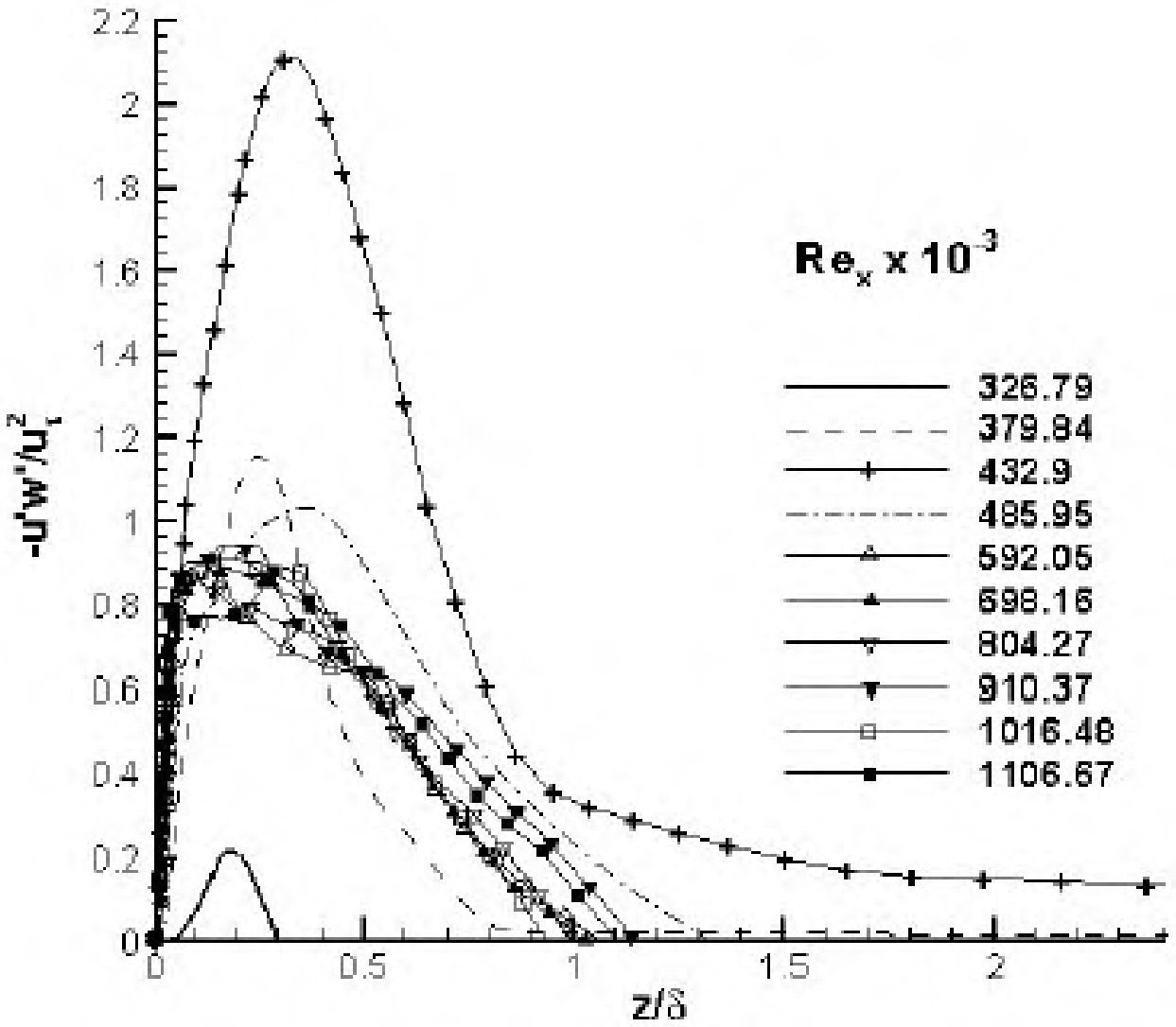

Figure 8: Reynolds stress

### 2.2.5 Comparison with Other DNS

Although we cannot compare our DNS results with those given by Borodulin et al (2002) quantitatively, we still can find that the shear layer structure are very similar in two DNS computations in Figure 9.

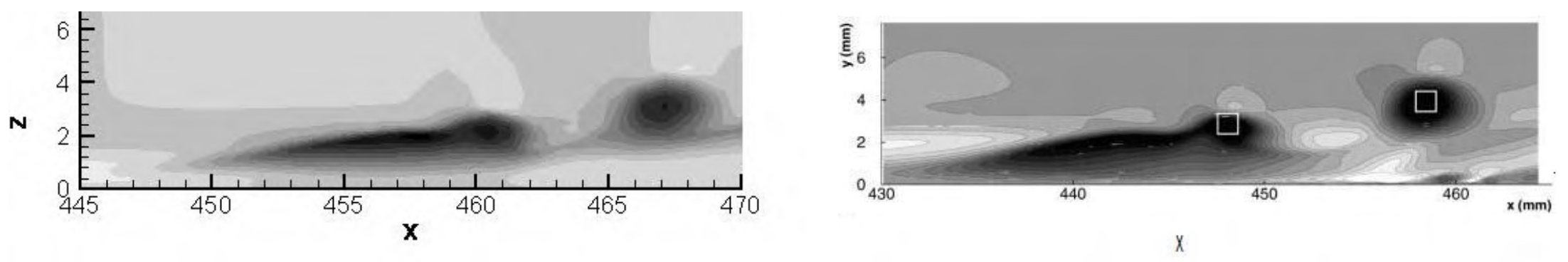

(a) Our DNS (b) Borodulin et al (2002)

Figure 9: Qualitatively Comparison of contours of streamwise velocity disturbance u in the (x, z)-plane (Light shades of gray correspond to high values)

### 2.2.6 Comparison with Experiment

By this $\Lambda_2$-eigenvalue visualization method, the vortex structures shaped by the nonlinear evolution of T-S waves in the transition process are shown in Figure 10. The evolution details are briefly studied in our previous paper (Chen et al 2009) and the formation of ring-like vortices chains is consistent with the experimental work (Lee C B & Li R Q, 2007, Figure 11) and previous numerical simulation by Rist and his co-authors (Bake et al 2002).

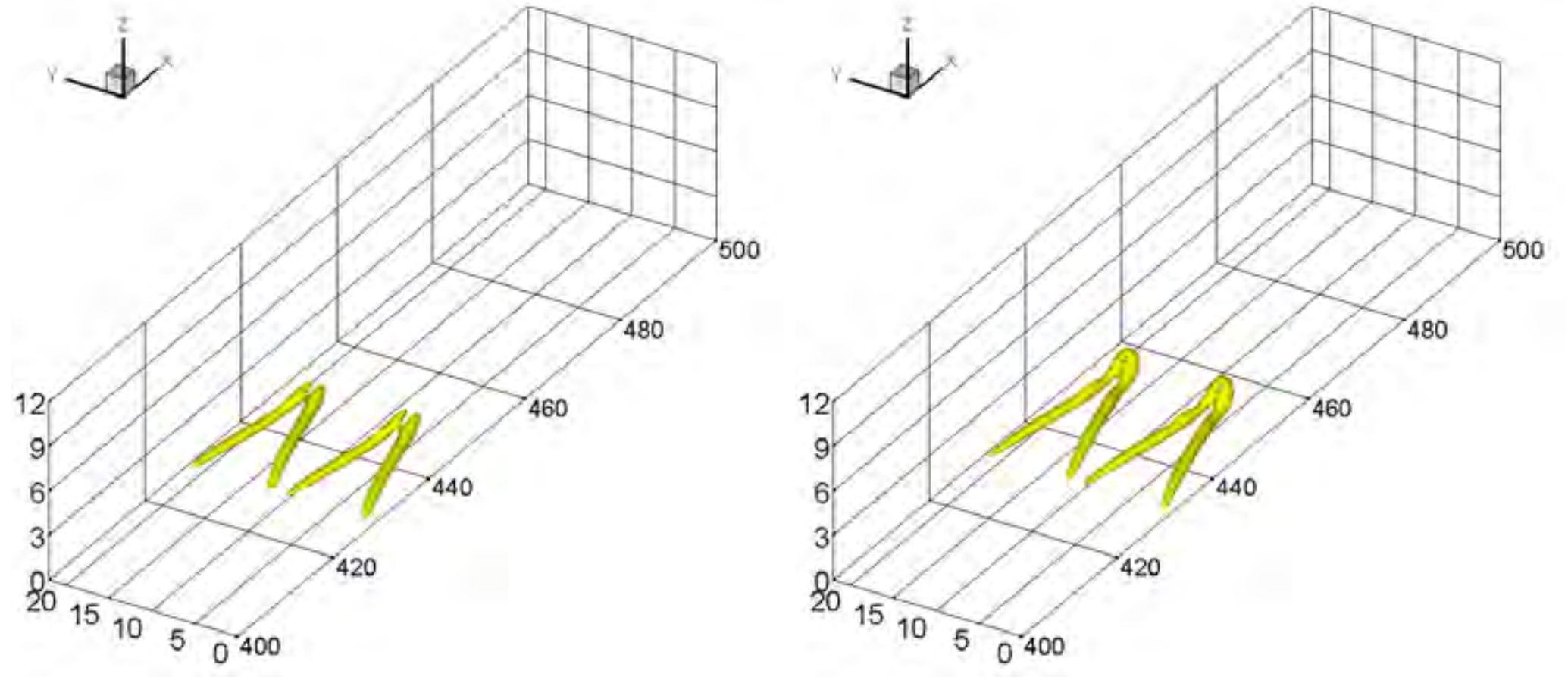

(a) $t = 6T$ (b) $t = 6.2T$

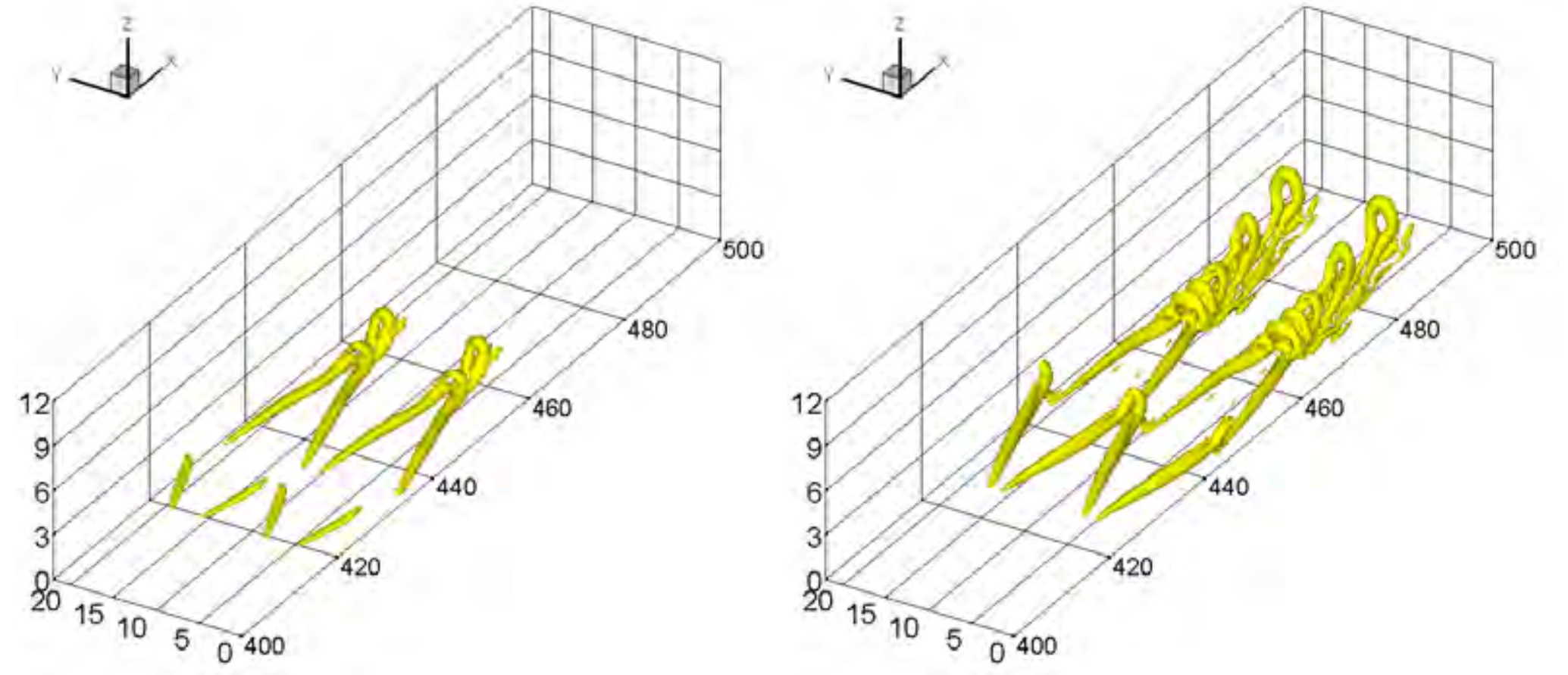

(c) $t = 6.4T$ (d) $t = 7.0T$

Figure 10: Evolution of vortex structure at the late-stage of transition (Where $T$ is the period of T-S wave)

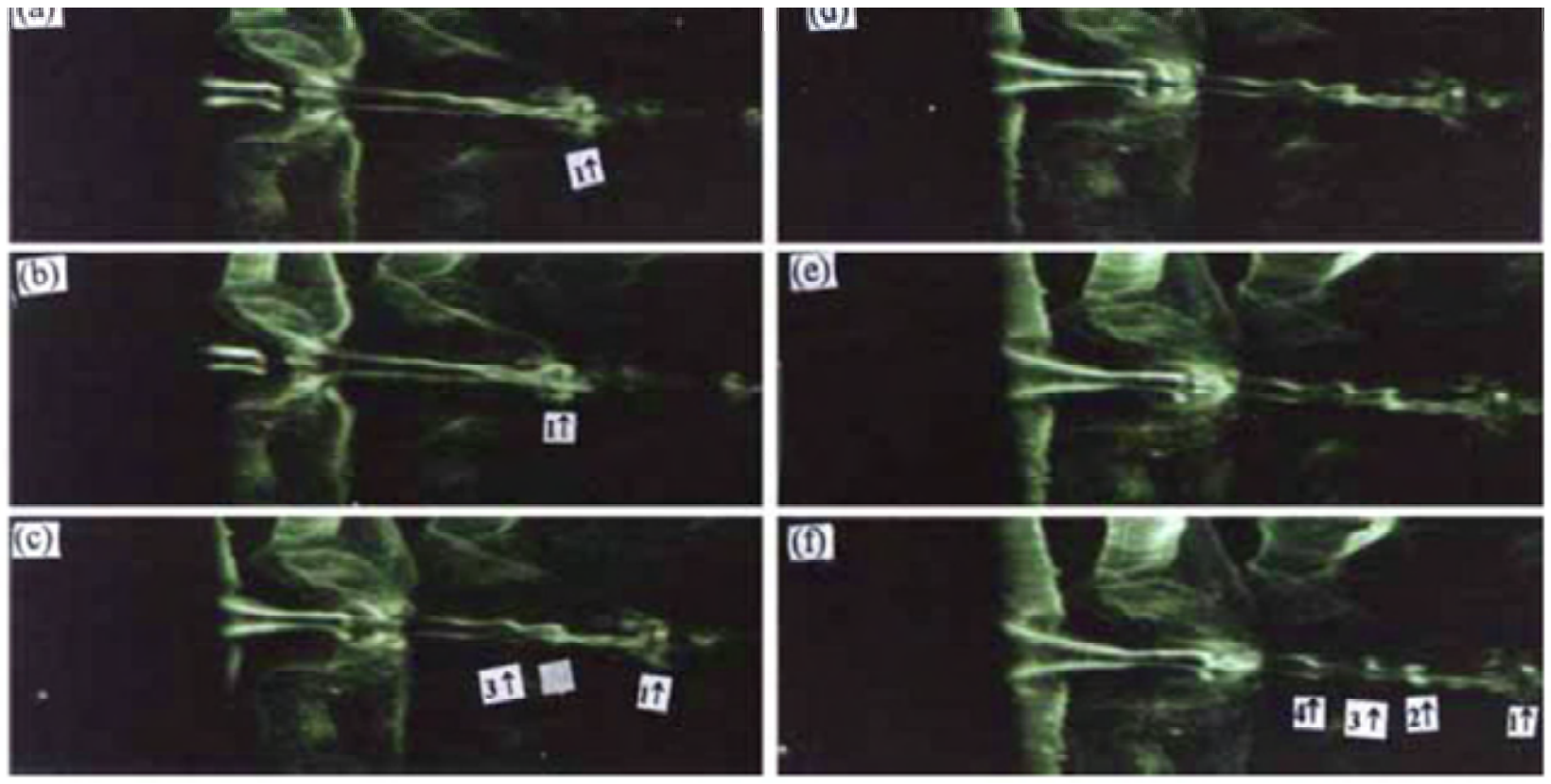

Figure 11: Evolution of the ring-like vortex chain by experiment (Lee et al, 2007)

### 2.2.7 U-shaped vortex in comparison with experimental results

Figure 12(a) (Guo et al, 2010) represents an experimental investigation of the vortex structure including ring-like vortex and barrel-shaped head (U-shaped vortex). The vortex structures of the nonlinear evolution of T-S waves in the transition process are given by DNS in Figure 12(b). By careful comparison between the experimental work and DNS, we note that the experiment and DNS agree with each other in a detailed flow structure comparison. **This cannot be obtained by accident, but provides the following clues: 1) Both DNS and experiment are correct 2) Disregarding the differences in inflow boundary conditions (random noises VS enforced T-S waves) and spanwise boundary conditions (non-periodic VS periodic) between experiment and DNS, the vortex structures are same 3) No matter K-, H- or other types of transition, the final vortex structures are same 4) There is a universal structure for late boundary layer transition 5) turbulence has certain coherent structures (CS) for generation and sustenance.**

**Coherent structure of U-shaped vortex**

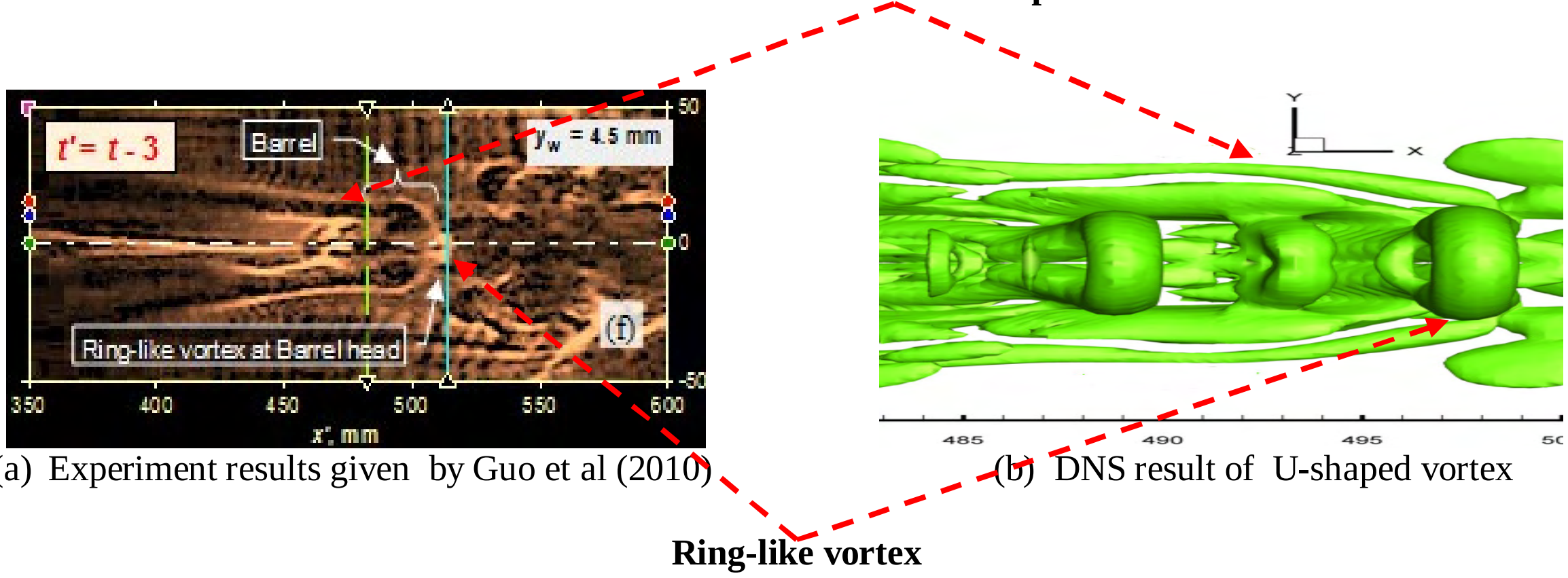

(a) Experiment results given by Guo et al (2010) (b) DNS result of U-shaped vortex

**Ring-like vortex**

Figure 12: Qualitative vortex structure comparison with experiment

**All these verifications and validations above show that our code is correct and our DNS results are reliable.**

## III. Observation and Analysis by Our DNS

**Following observations have been made and reported by our previous publications** (Chen et al., 2009, 2010a, 2010b, 2011a, 2011b; Liu et al., 2010a, 2010b, 2010c, 2011a, 2011b 2011c, 2011d; Lu et al., 2011a, 2011b).

### 3.1 Mechanism of large coherent vortex structure

Late boundary transition starts from the formation of the first vortex ring. Ring-like vortices play a vital role in the transition process of the boundary layer flow and must be studied carefully.

#### 3.1.1 Vortex rings in flow field (Chen et al 2011a)

According to Helmholtz vorticity conservation law, the vortex tube cannot have ends inside the flow field. The only form of vortex tube existing inside the flow filed must be ring with or without legs (Figure 13)

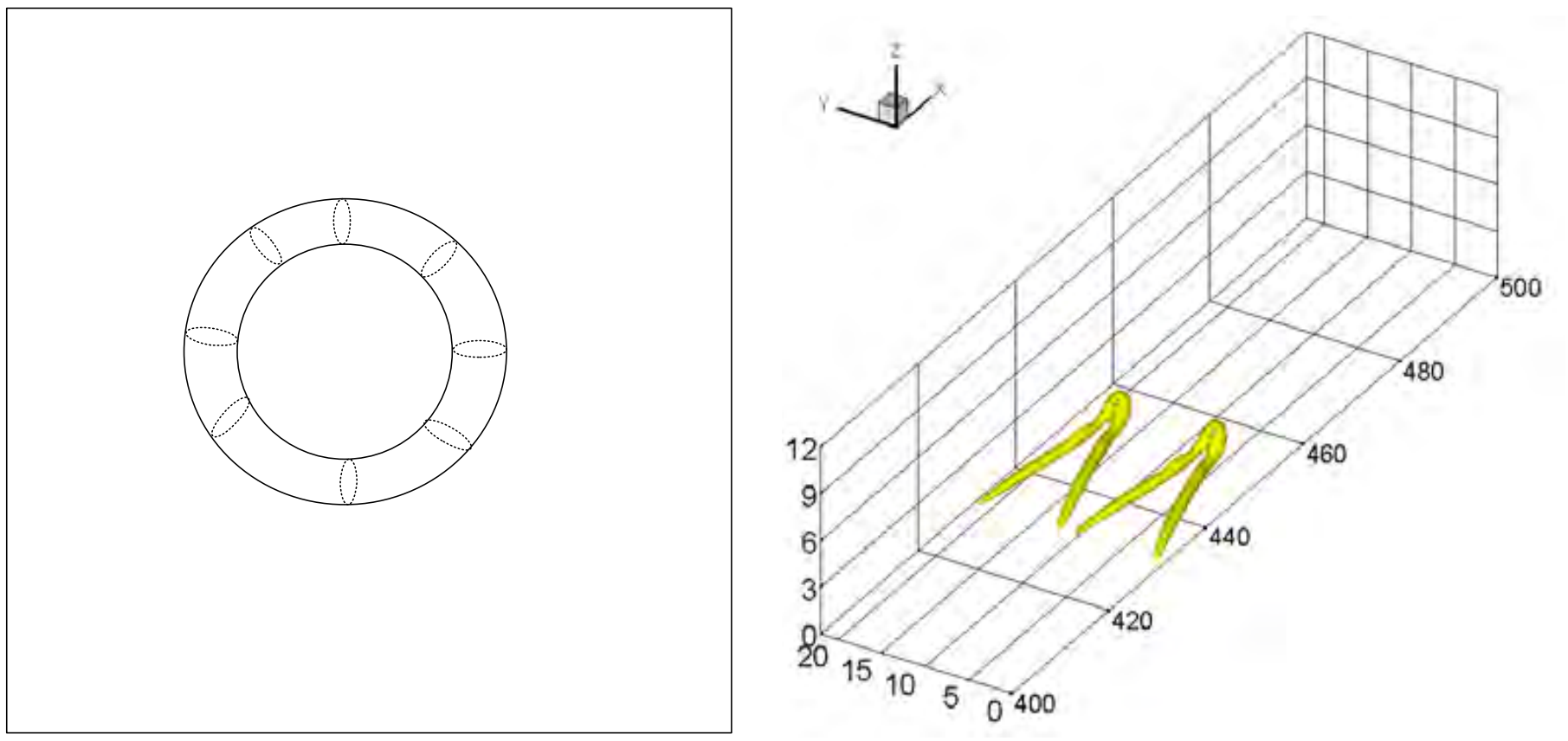

(a) Vortex ring without legs (b) Vortex ring with legs

Figure 13: Vortex rings with or without legs

#### 3.1.2 Mechanism of first ring-like vortex formation (Chen et al 2011a)

The ring shape of first ring-like vortex is caused by the interaction between the prime streamwise vortices and secondary streamwise vortices (Figure 14), which is quite different from one given by Moin et al (1986). The ring-like vortex is located at the edge of the boundary layer or in other words located in the inviscid zone ( $z = 3.56, U = 0.99Ue$ ). The vortex ring is perpendicularly standing and the shape of the ring is almost perfectly circular, which will generate the strongest downwash sweeps (Figures 14 and 15). More important, no vortex "pinch off" is observed.

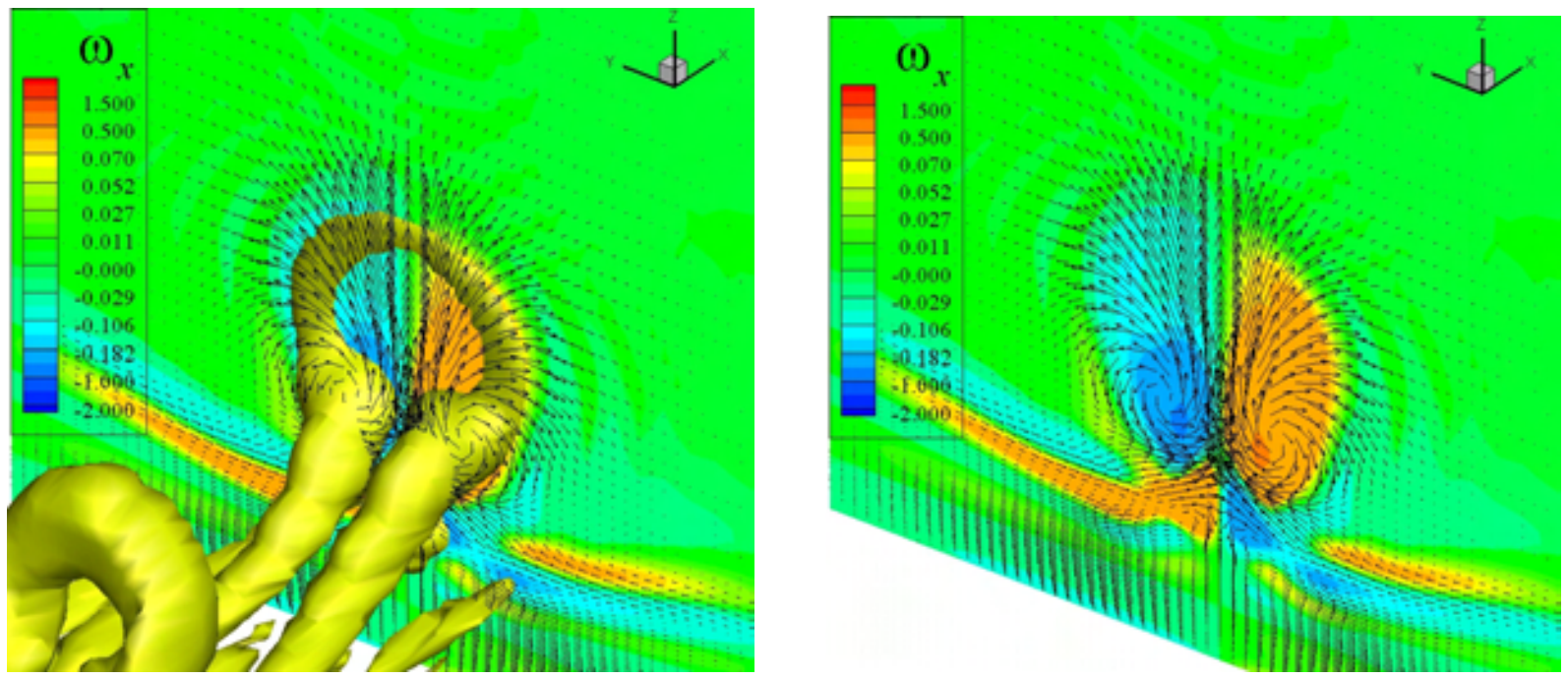

Figure 14: Formation processes of first ring-like vortex

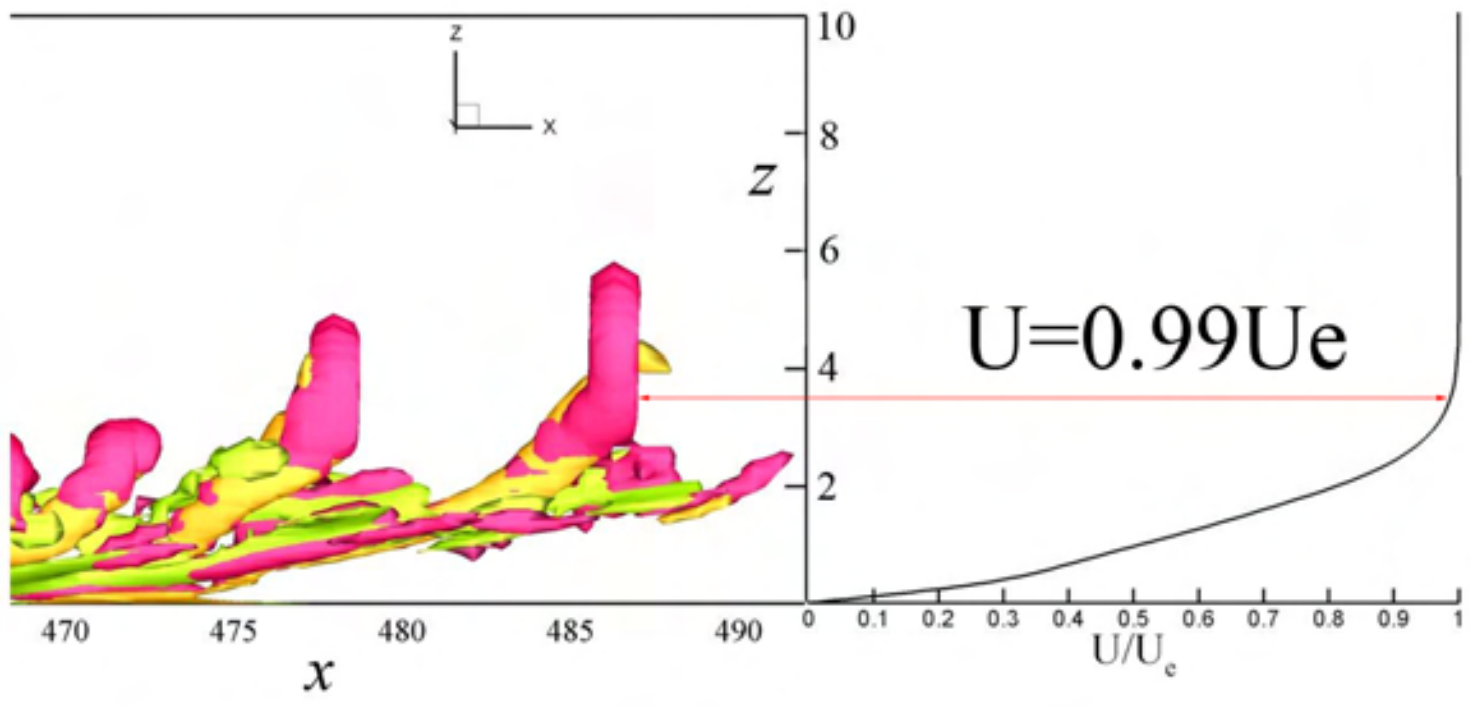

Figure 15: The shape and position of ring-like vortices in boundary layer ( $z = 3.56, U = 0.99Ue$ )

### 3.1.3 Mechanism of multiple ring formation *(new)*

The ejection by rotation of the primary vortex legs brings low speed flow from lower boundary layer up and forms a cylinder-like momentum deficit zone in the middle of the two legs (green in Figure 16). The momentum deficit zone is also located above the legs and forms shear layers due to the stream-wise velocity difference between surrounding high speed flow and low speed flow in the deficit zone. These shear layers are not stable and multiple ring-like vortices (yellow in Figure 16) are generated by the shear layers one by one (Figure 17). This process must satisfy the Helmholtz vorticity conservation and the primary streamwise vorticity must be reduced when a new spanwise-oriented ring is formed (Figure 18). Different from Borodulin et al (2002), neither “vortex breakdown and reconnection” nor ”Crow theory” is observed by our new DNS.

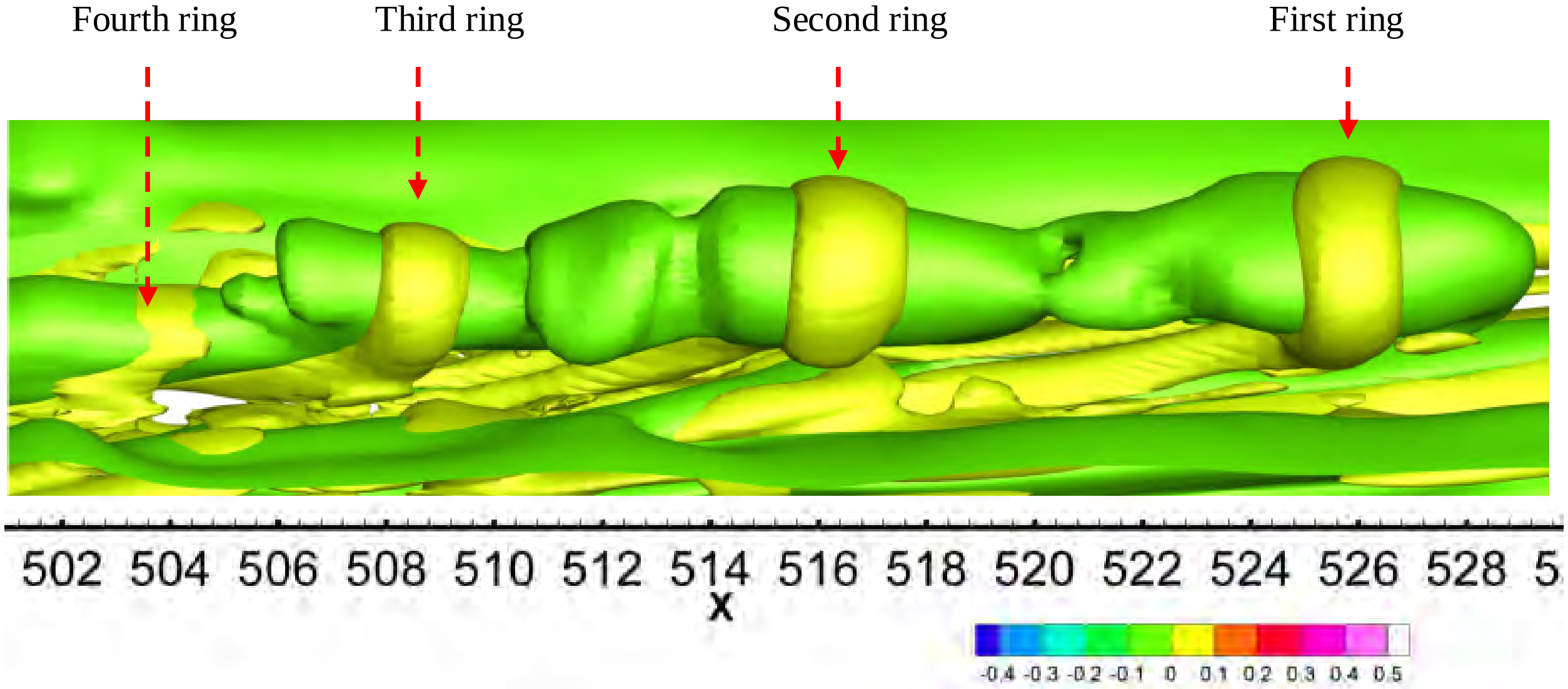

Figure 16: Three-dimensional momentum deficit and multiple ring-like vortex formation

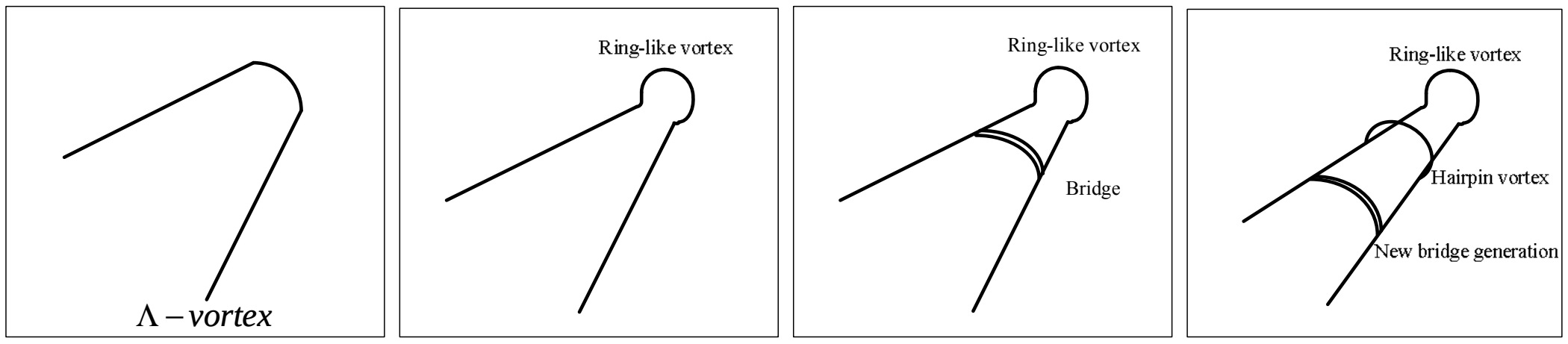

Figure 17 (a): Sketch for mechanism of multi-ring generation

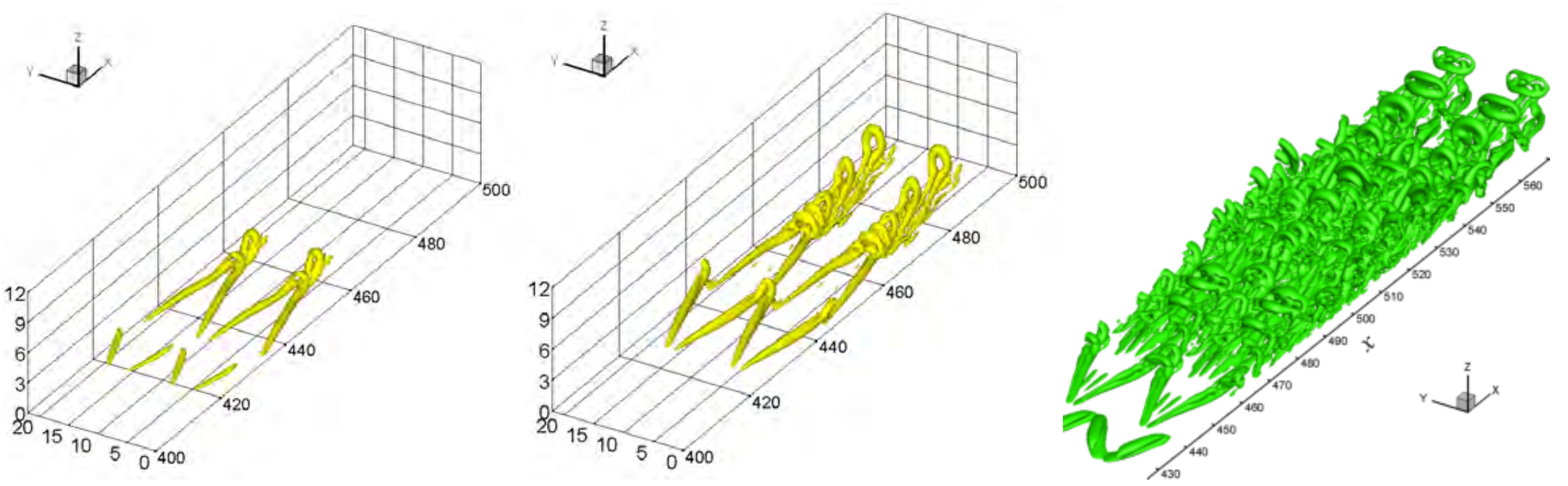

Figure 17 (b): Multiple ring formation

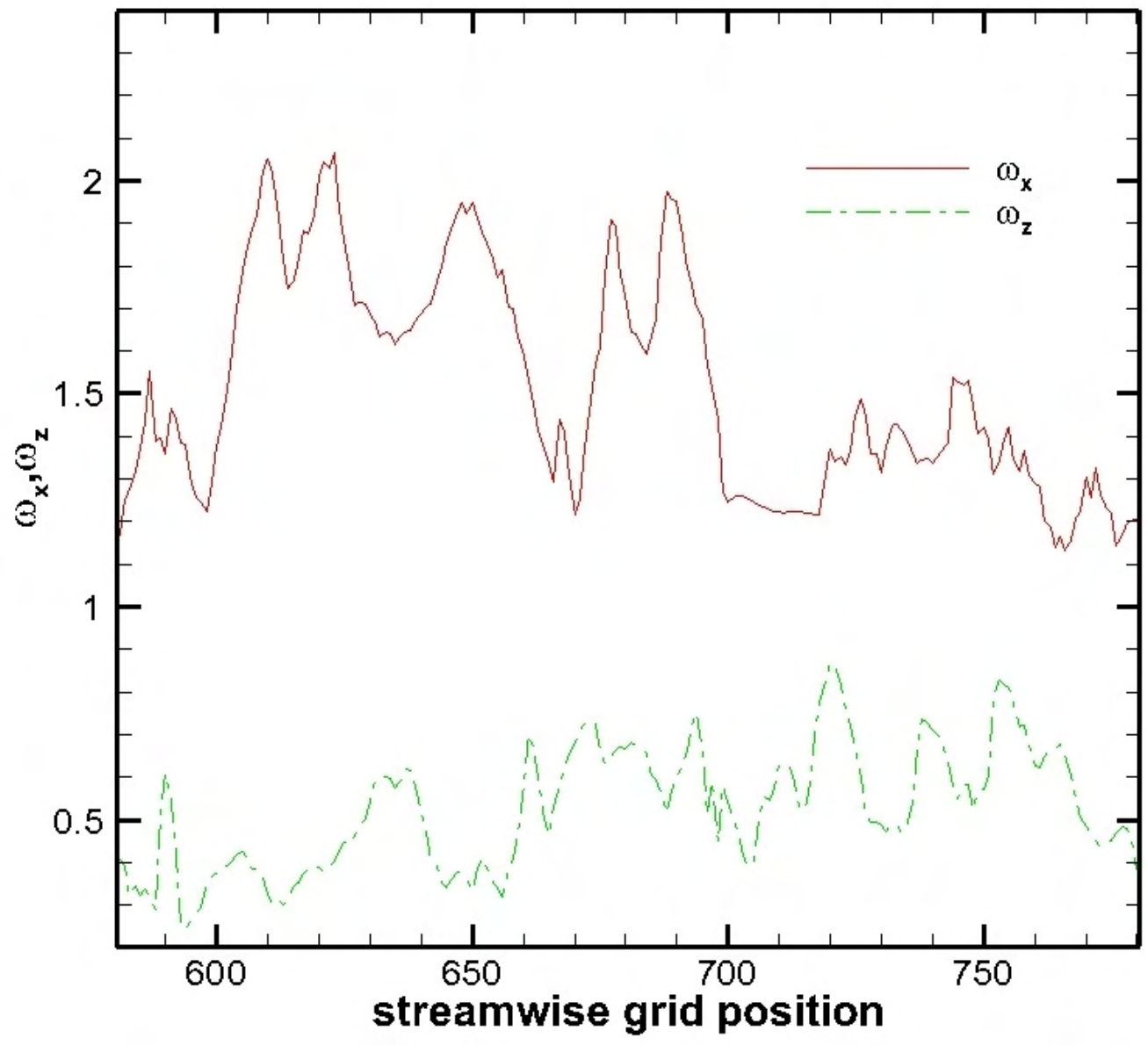

Figure 18: Maximum streamwise and spanwise vorticity along x-coordinate (Helmoholtz vorticity conservation)

### 3.1.4 **Primary, Secondary and U-shaped vortex** (Lu et al, 2011a)

U-shaped vortex is part of large coherent vortex structure and is generated by secondary vortices which are induced by the primary vortex. Being different from Guo et el (2010), the U-shaped vortex is not a wave and is not induced by second sweeps and positive spikes. Actually, the U-shaped vortex is a tertiary (not secondary) vortex with same vorticity sign as the original ring legs (Figures 19 and 20). In addition, the U-shaped vortices serve as additional channels to supply vorticity to the multiple ring structure ( Lu et al 2011b).

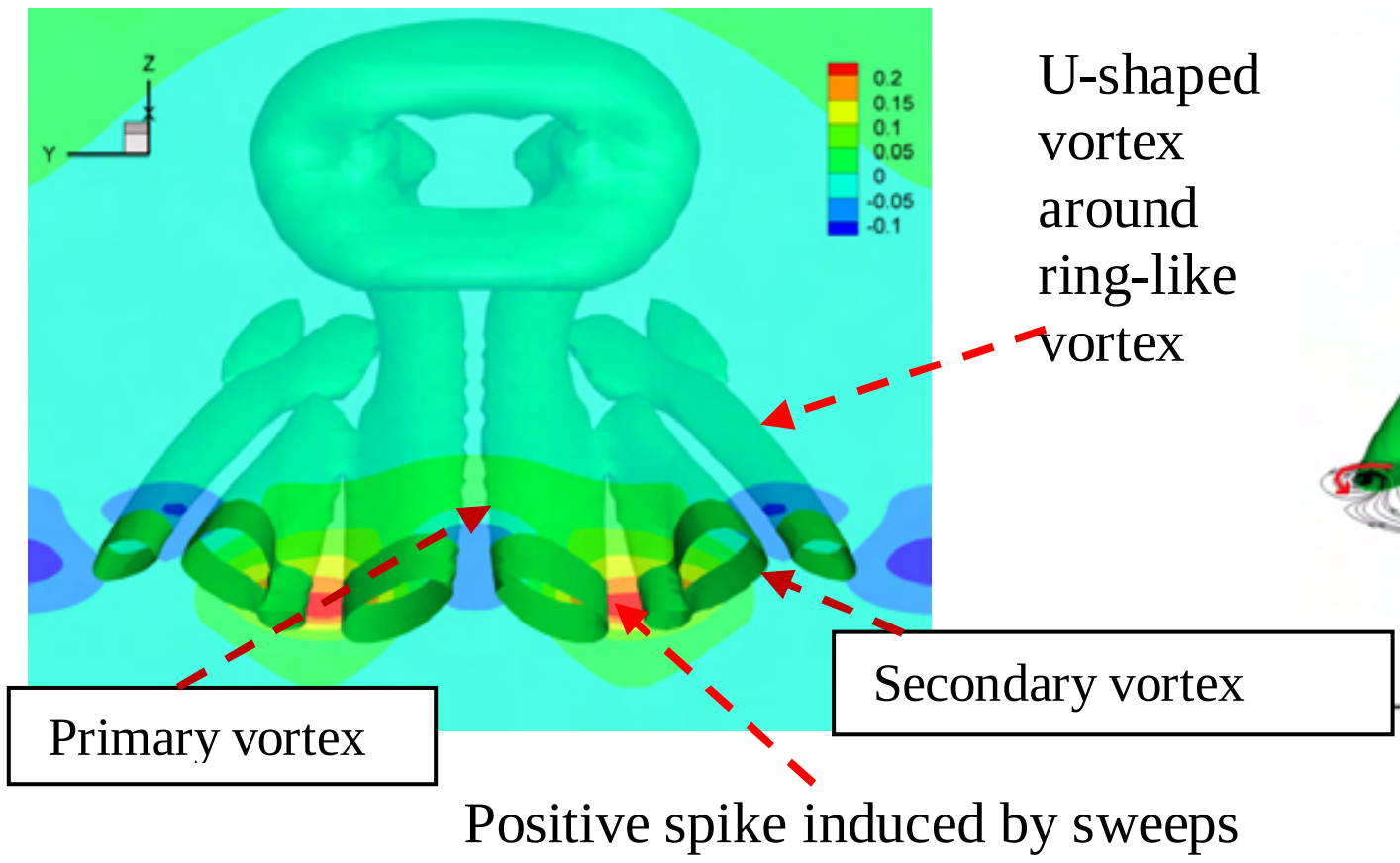

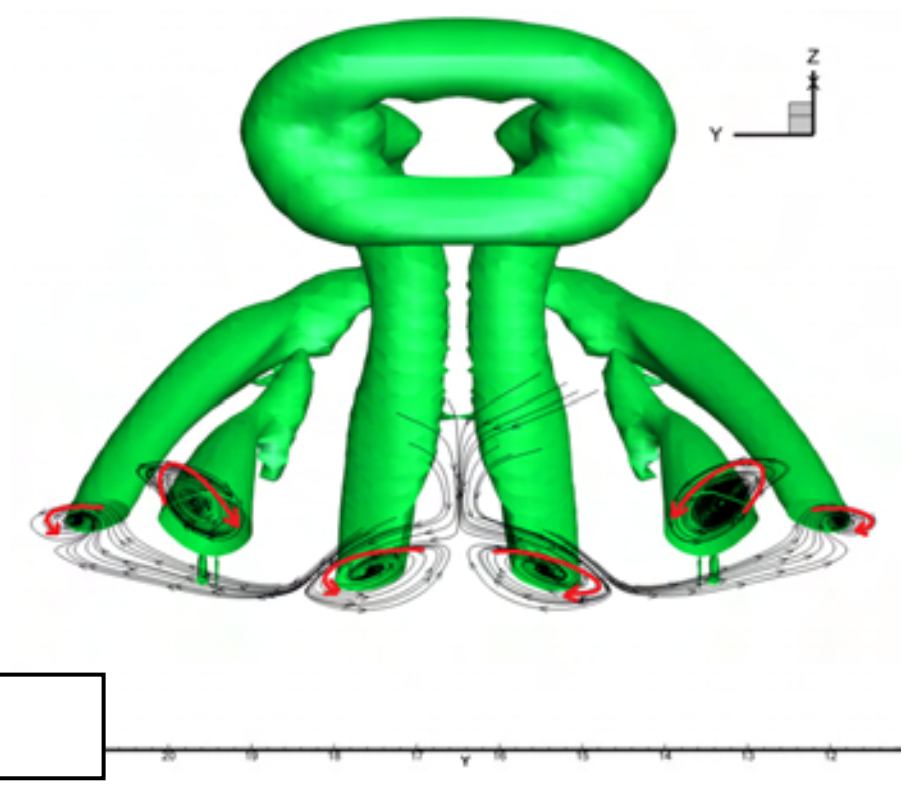

Figure 19: U-shaped vortex with streamwise velocity perturbation contour

Figure 20: Isosurface of $\lambda_2$ and streamtrace at x=530.348 $_{in}$

**3.2Mechanism of small length vortices generation**

Turbulence has two futures: 1) Small length vortices; 2) Non-symmetric structure (Randomization)

3.2.1 **"Vortex breakdown to turbulence" is challenged** (Liu et al, 2011c, 2011d)

In the new DNS (Liu et al, 2011a, 2011b), it has been observed that the multiple vortex structure is a quite stable structure and never breaks down. Previously reported "vortex breakdown" is either based on 2-D visualization or made by using low pressure center as the vortex center (Figure 21(a)). We can use a different $\Lambda_2$ value to get similar "vortex breakdown" (Figure 21(b)), which is faked. Here, we define vortex as a tube with a rotated core and iso-sufaces of constant vorticity flux. However, there is no evidence that the vortex breaks down. Let us look at the head of the so-called "turbulence spot" from different directions of view (Figures 22). Due to the increase of the ring (bridge) number and vorticity conservation, the leading rings will become weaker and weaker until they cannot be detected, but they never break down (see Figure 23).

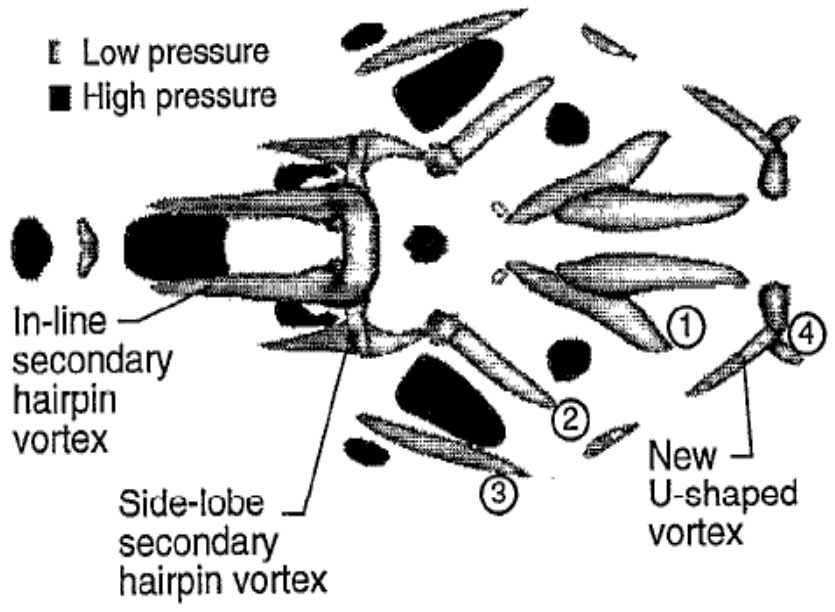

Figure 21(a): U-shaped vortex (Singer, 1994)

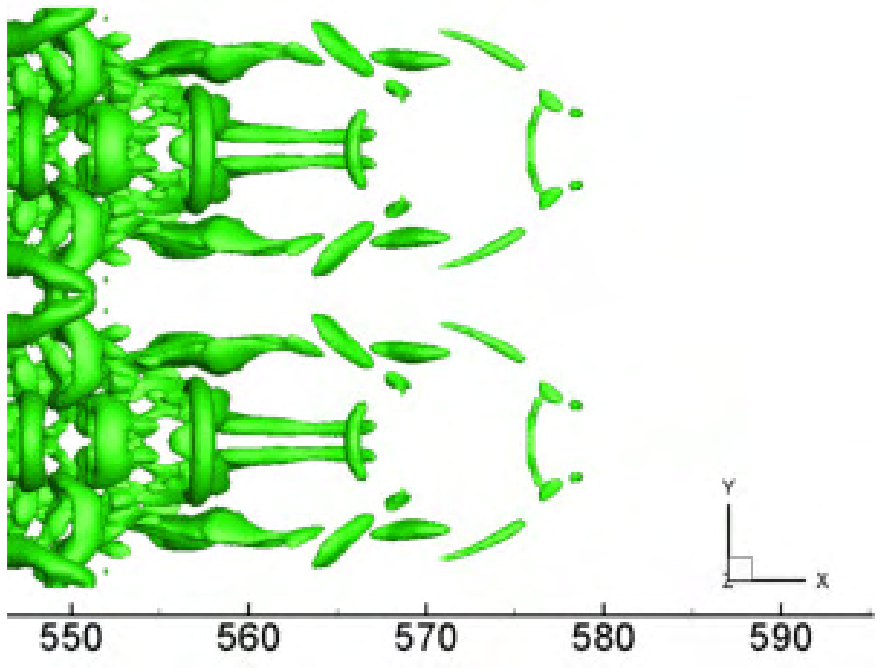

Figure 21(b): DNS with a different λ-value (fake)

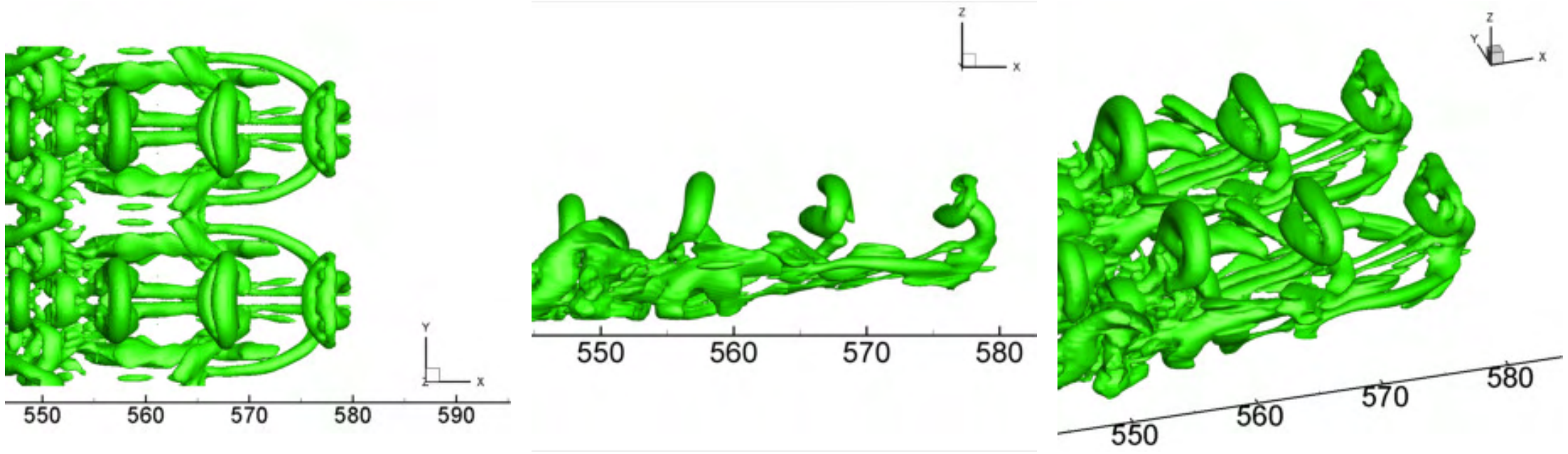

(a) Top view (b) Side view (c) Angle view

Figure 22 View of young turbulence spot head from different directions (t=8.8T)

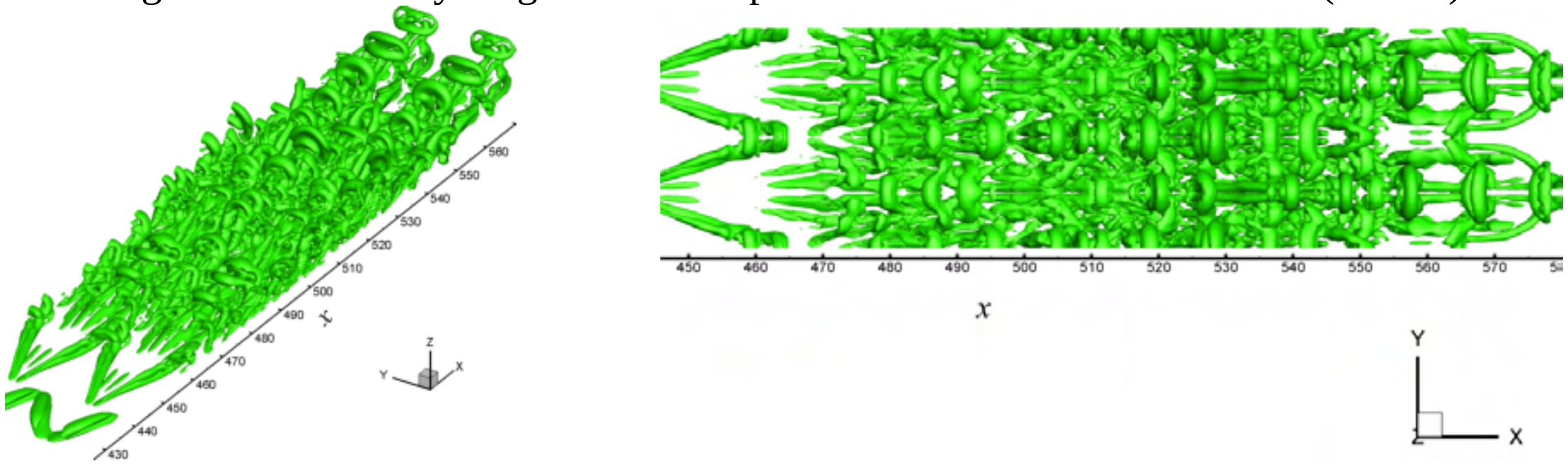

(a) Angle view (b) Top view

Figure 23 3-D angle view and 2-D top view of the young turbulence spot (t=8.8T)

**3.2.2 Mechanism of small length scale (turbulence) generation – by "shear layers"** (Liu et al, 2011d; Lu et al, 2011b)

The question will be raised that where the small length scales come from if the small vorticies are not generated by "vortex breakdown"? Since we believe that the small vortices are generated by shear layers near the wall surface, we take snap shots in the direction of view from the bottom to top (Figure 24). The evidence provided by our new DNS confirms that the small length scale vortices are generated near the bottom of the boundary layer. Being different from classical theory which believes turbulence is generated by "vortex breakdown", our new DNS found that all small length scales (turbulence) are generated by high shear layers (HS) near the bottom of the boundary layers (near the laminar sub-layers). There is no exception. When we look at the later stage of flow transition at t=15T, we can see that all small length scales are located around the high-shear layers, especially at those near the wall surface (Figures 25-26).

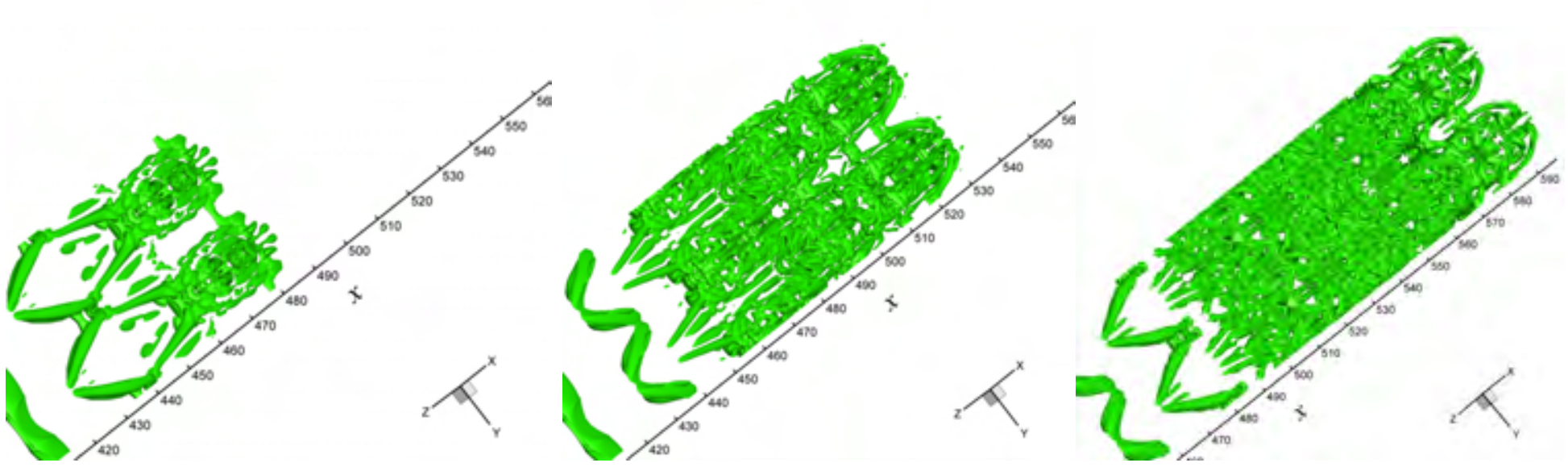

(a) $t = 7.0T$ (b) t=8.0T (c) t=8.8T

Figure 24: Small length scale vortex generation at different time steps (view up from bottom): small length scale vortices are generated by the solid wall near the ring necks from the beginning to the end

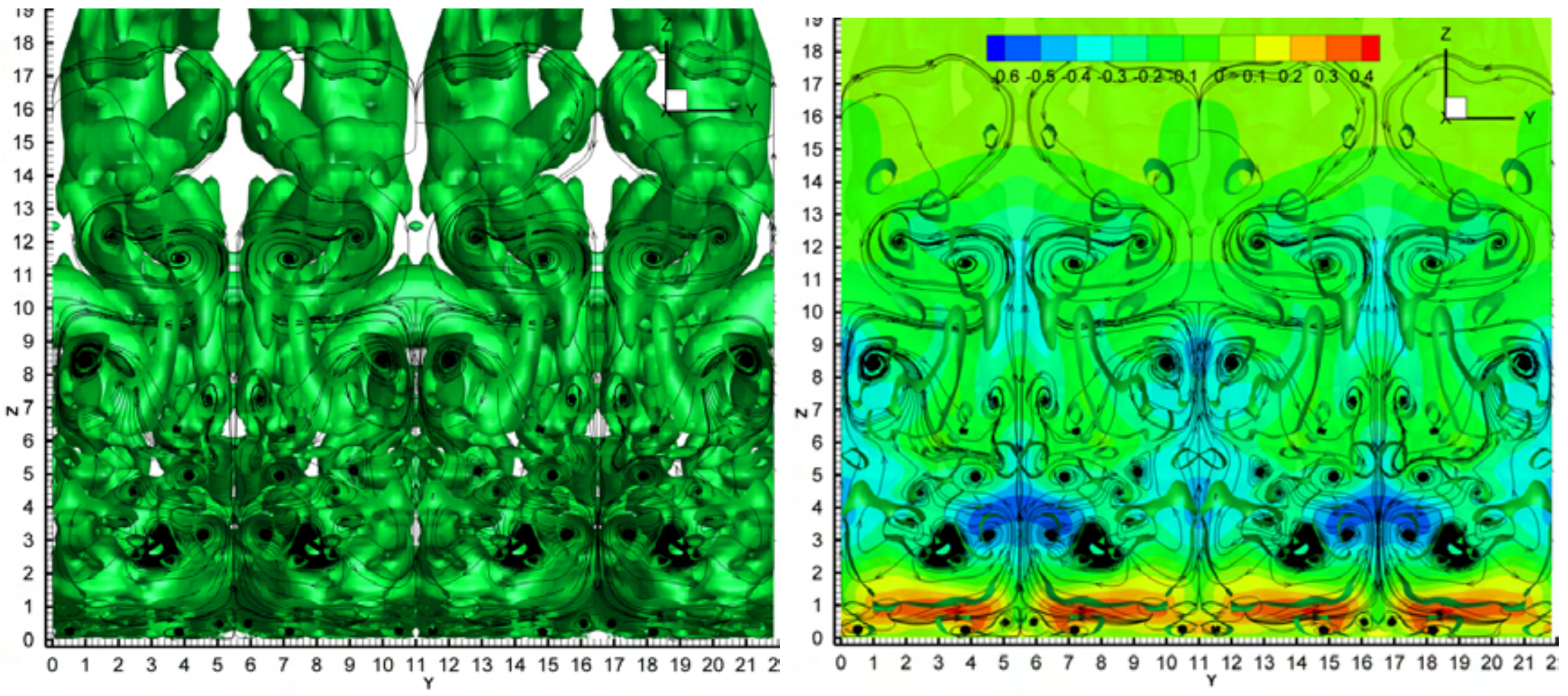

(a) $\Lambda_2$ isosurfaces (b) Contour of velocity perturbation

Figure 25: Visualization of isosurface $\Lambda_2$ and velocity perturbation at x=508.633 for (Y, Z)-Plane t=15T

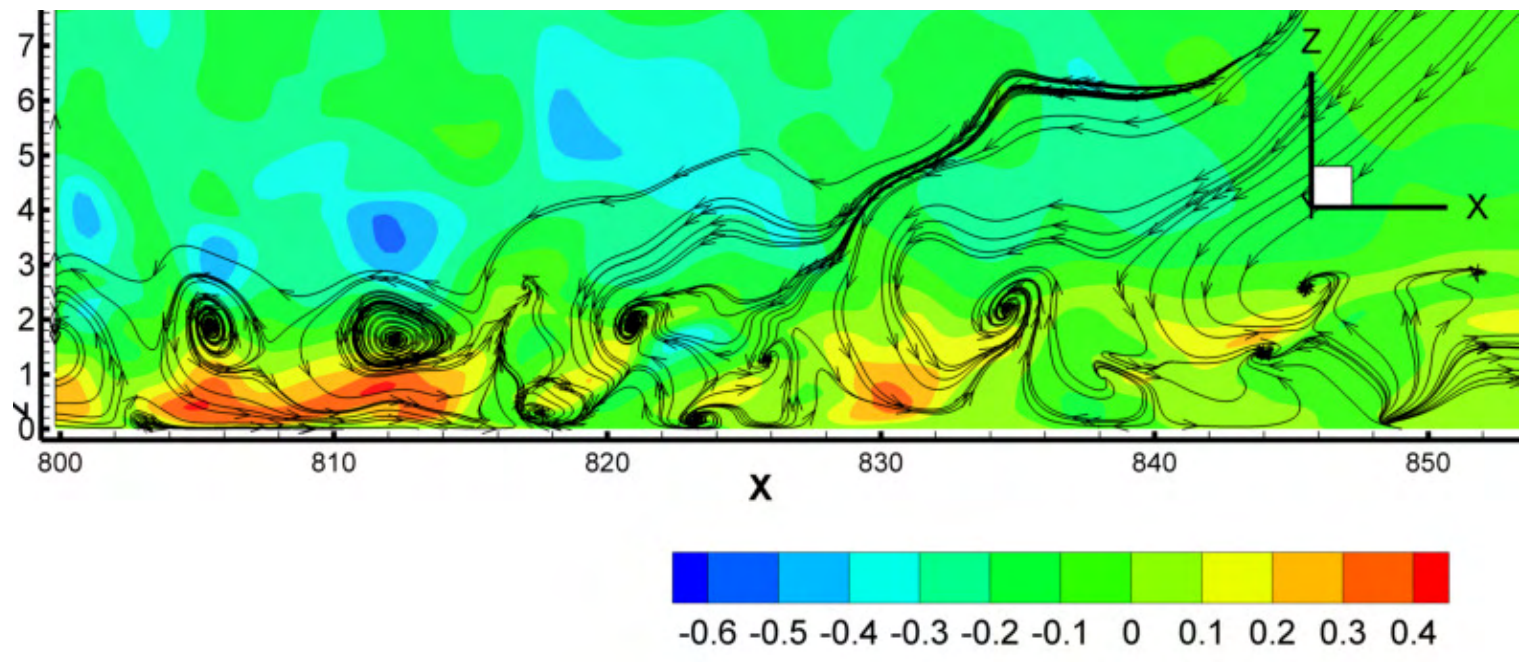

Figure 26: Visualization of isosurface $\Lambda_2$ and velocity perturbation at y=4 for (X, Z)-Plane t=15T

**3.2.3 High shear layer formation by sweeps** (Liu et al, 2011d)

Experiment and our numerical results have confirmed that there is a second sweep excited by every ring-like vortex (Figure 27). Combined with the first sweep generated by the original $\Lambda$vortex legs, it forms strong positive spikes which generate strong high shears at the bottom of the boundary layer (red in Figure 28). Figures 29 and 30 are contours of isosurface of $\Lambda_2$ and velocity perturbation at x=508.66. The second sweep movement induced by ring-like vortices working together with first sweep will lead to huge energy and momentum transformation from high energy inviscid zones to low energy zones near the bottom of the boundary layers and we can observe that **all small length scales are generated under the high speed region corresponding to high shear layers between the positive spike (momentum increment) and solid wall surface.**

**3.2.4 Multiple level sweeps and multiple level negative and positive spikes** (Lu et al, 2011b)

The positive spikes (momentum increment) could generate new ring-like vortices. The new ring-like vortices can further generate new sweeps and form new positive spikes (Figure 31) at the location which are very close to the bottom of the boundary layer (laminar sub-layer). The new positive spike could induce new smaller vortex rings by unstable shear layers. These multilevel sweeps and multilevel negative and positive spikes provide channels for energy transfer from the inviscid area (high energy area) down to the bottom (low energy area). **This is the mechanism why turbulence can be generated** **and sustained. This mechanism can be interpreted as a universal mechanism for both transitional and turbulent flow in a boundary layer.**

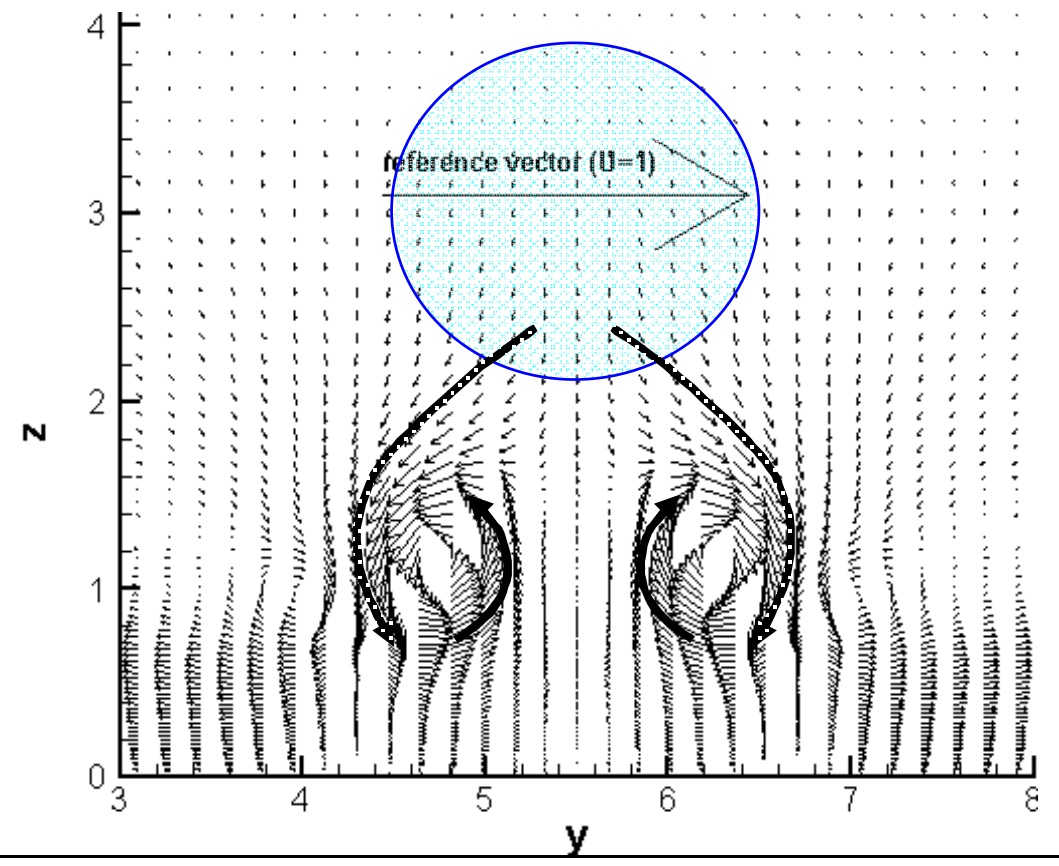

Figure 27: Velocity vector field

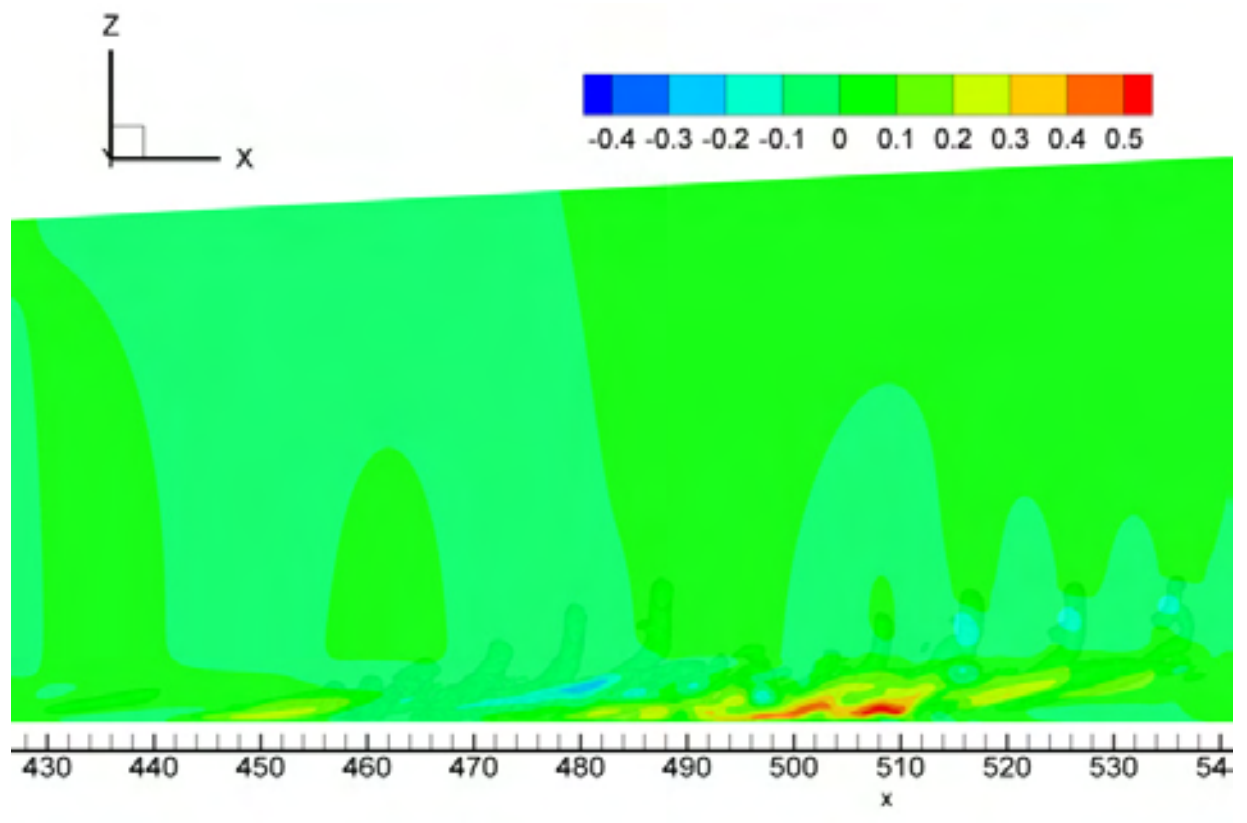

Figure 28: Vorticity distribution at x=508.633

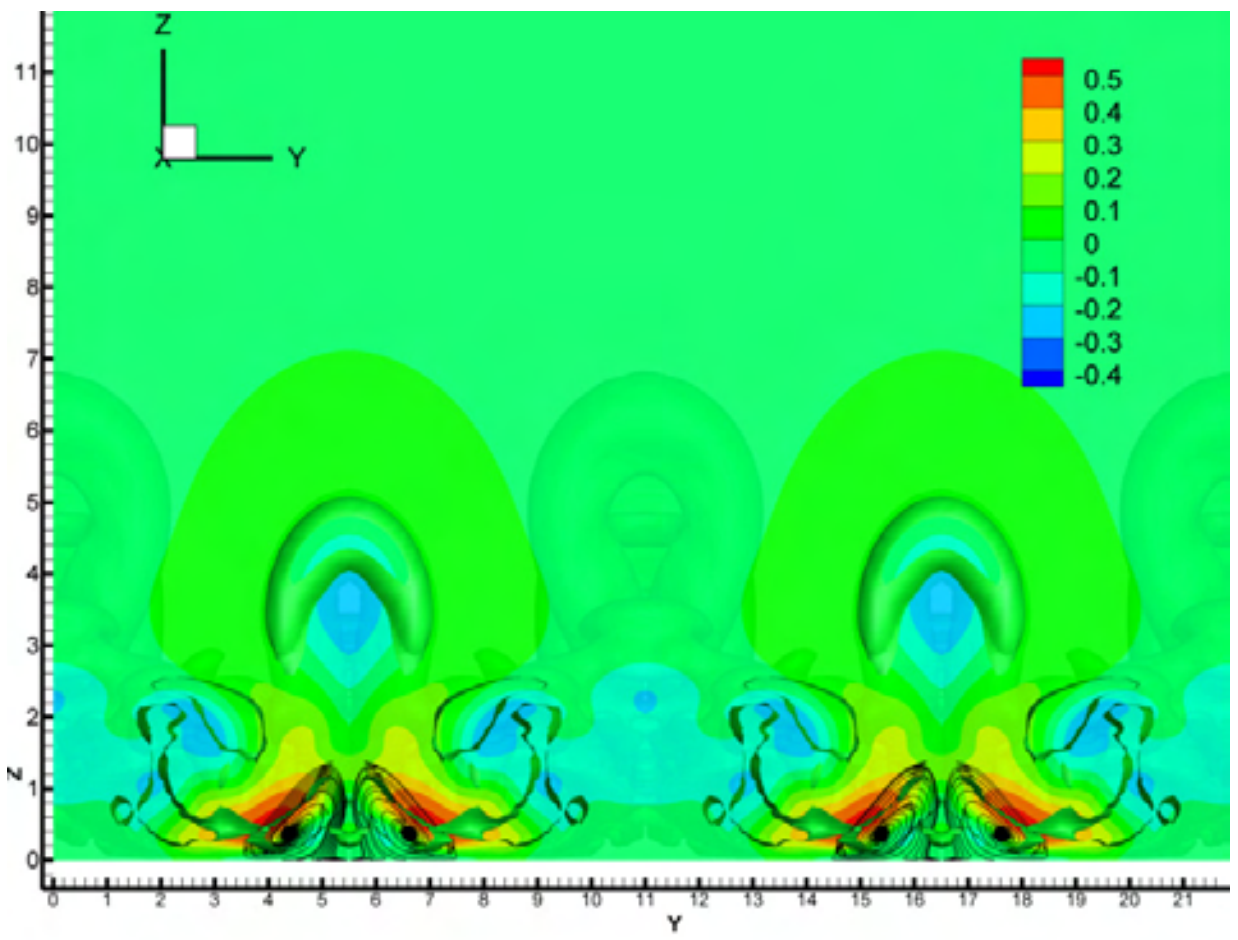

Figure 29: isosurface of $\Lambda_2$ and velocity at x=508.66

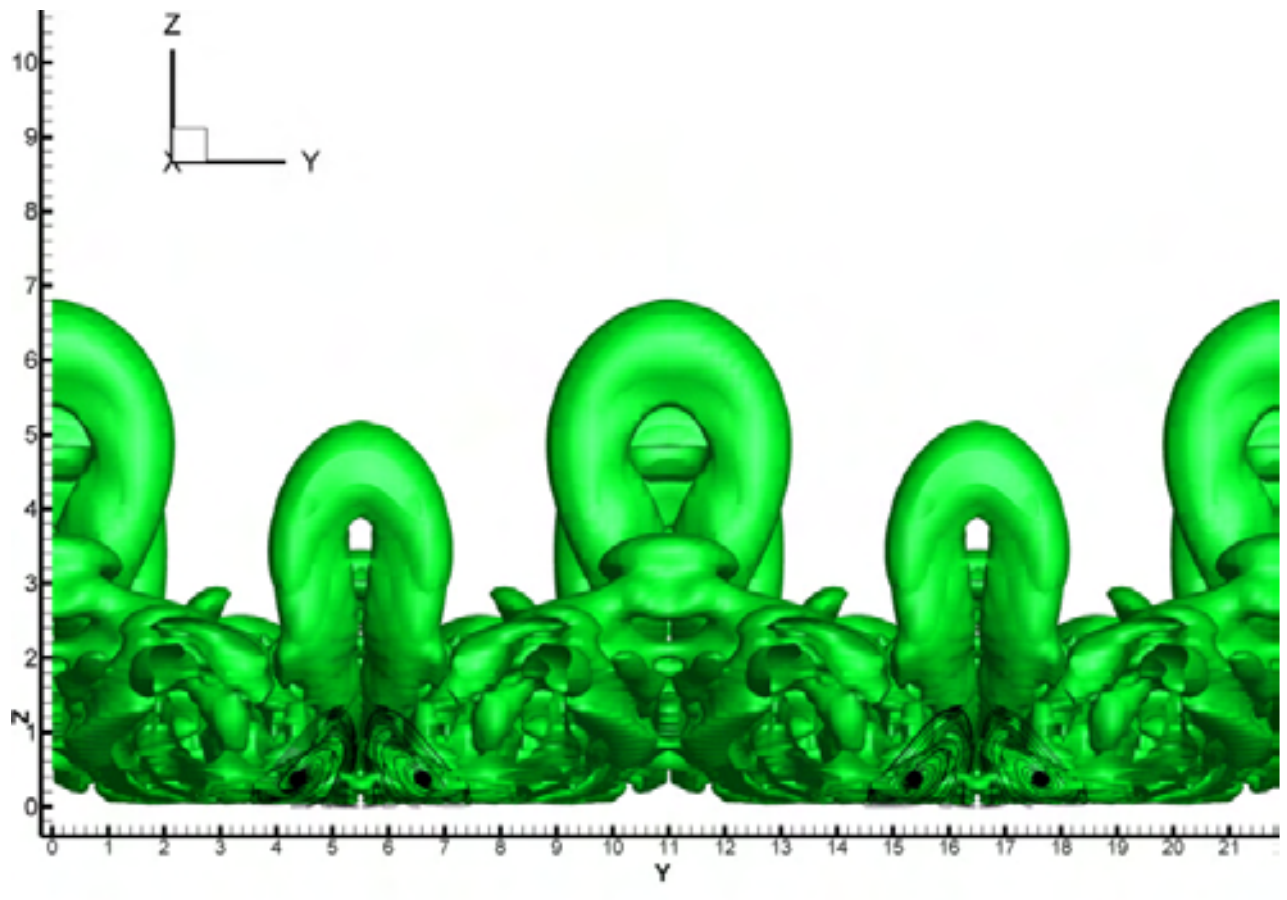

Figure 30: isosurface of $\Lambda_2$ and stream traces at x=508.66

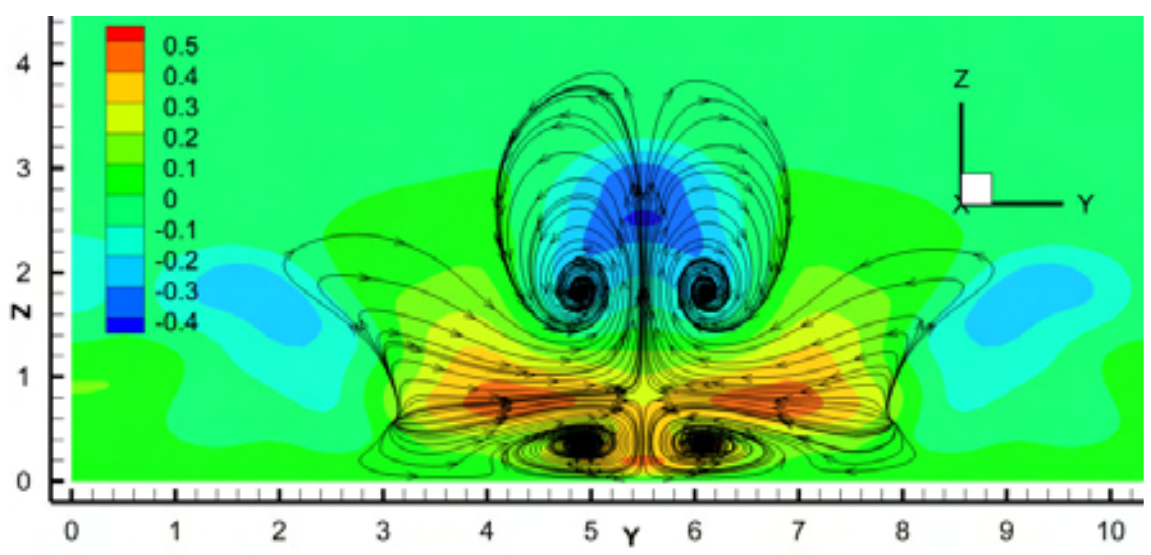

(a) Streamwise velocity perturbation and stream trace

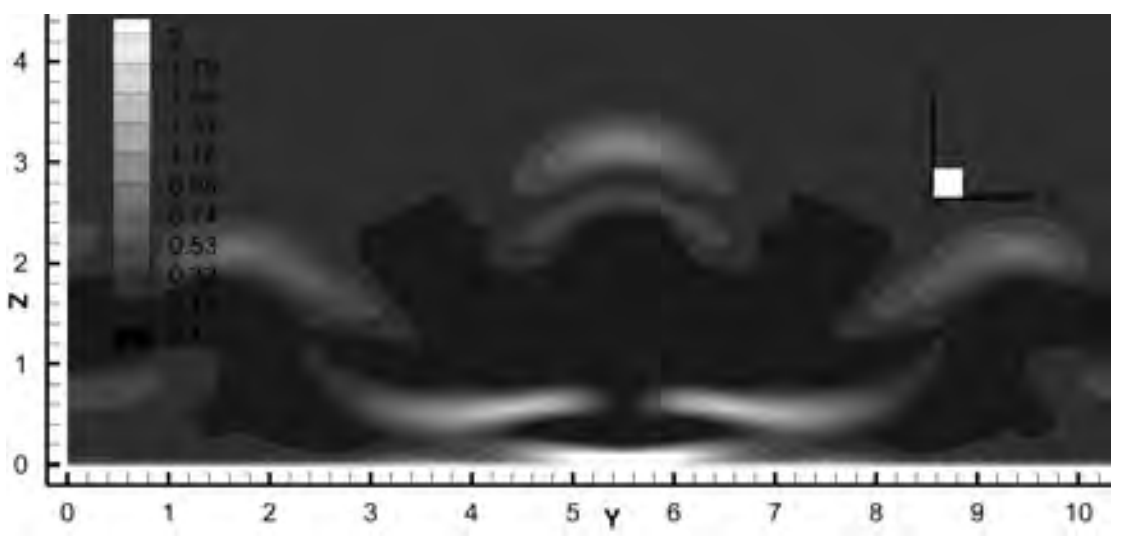

(b) Spanwise vorticity distribution

Figure 31: (a) Multilevel sweeps (b) multilevel positive spikes

**3.2.5 Energy transfer paths and universal turbulence spot structure** (Lu et al 2011b)

Figure 32 and 33 are sketches describing energy transfer and likely universal turbulence structure

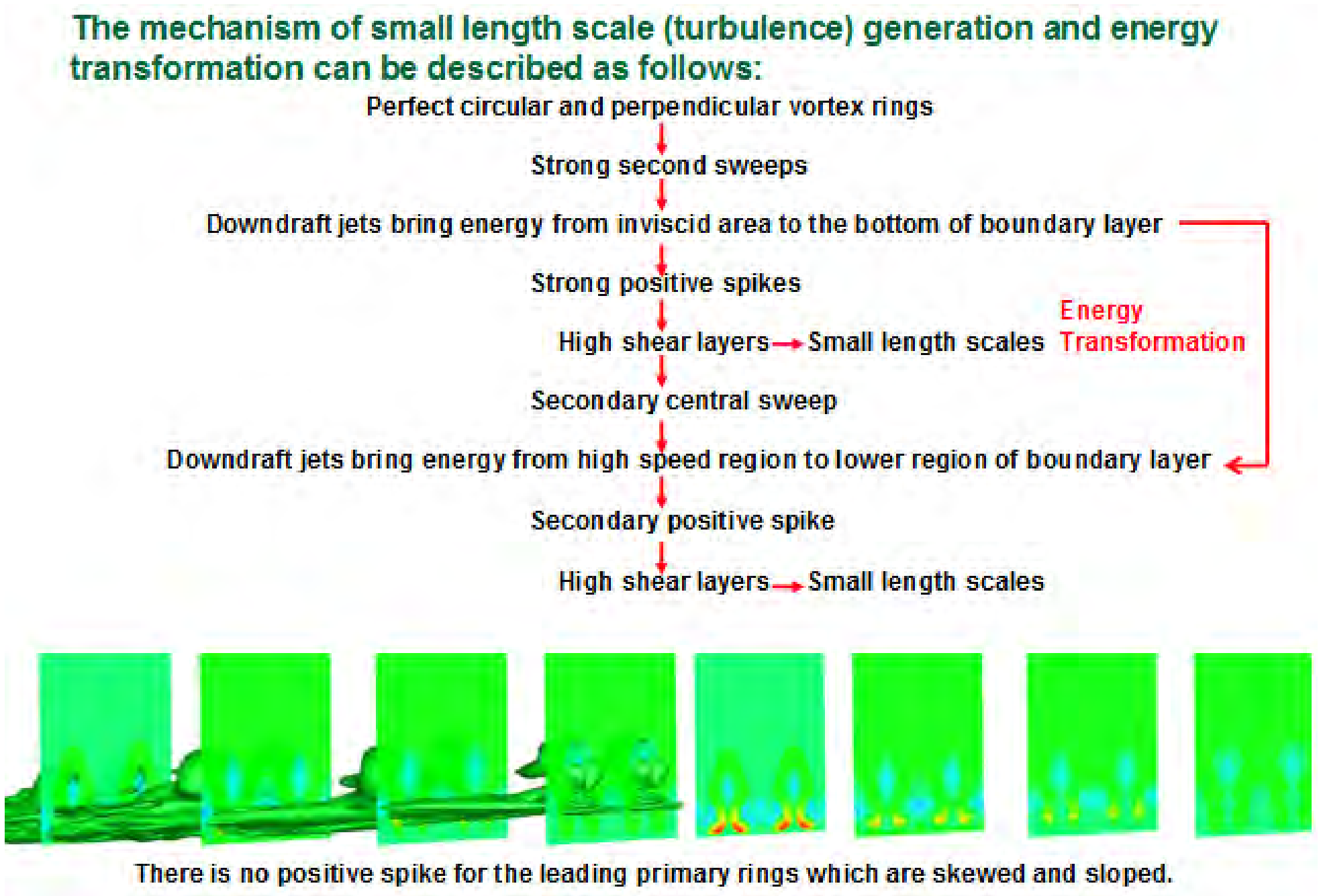

Figure 32: Energy transfer paths and universal structure for turbulence

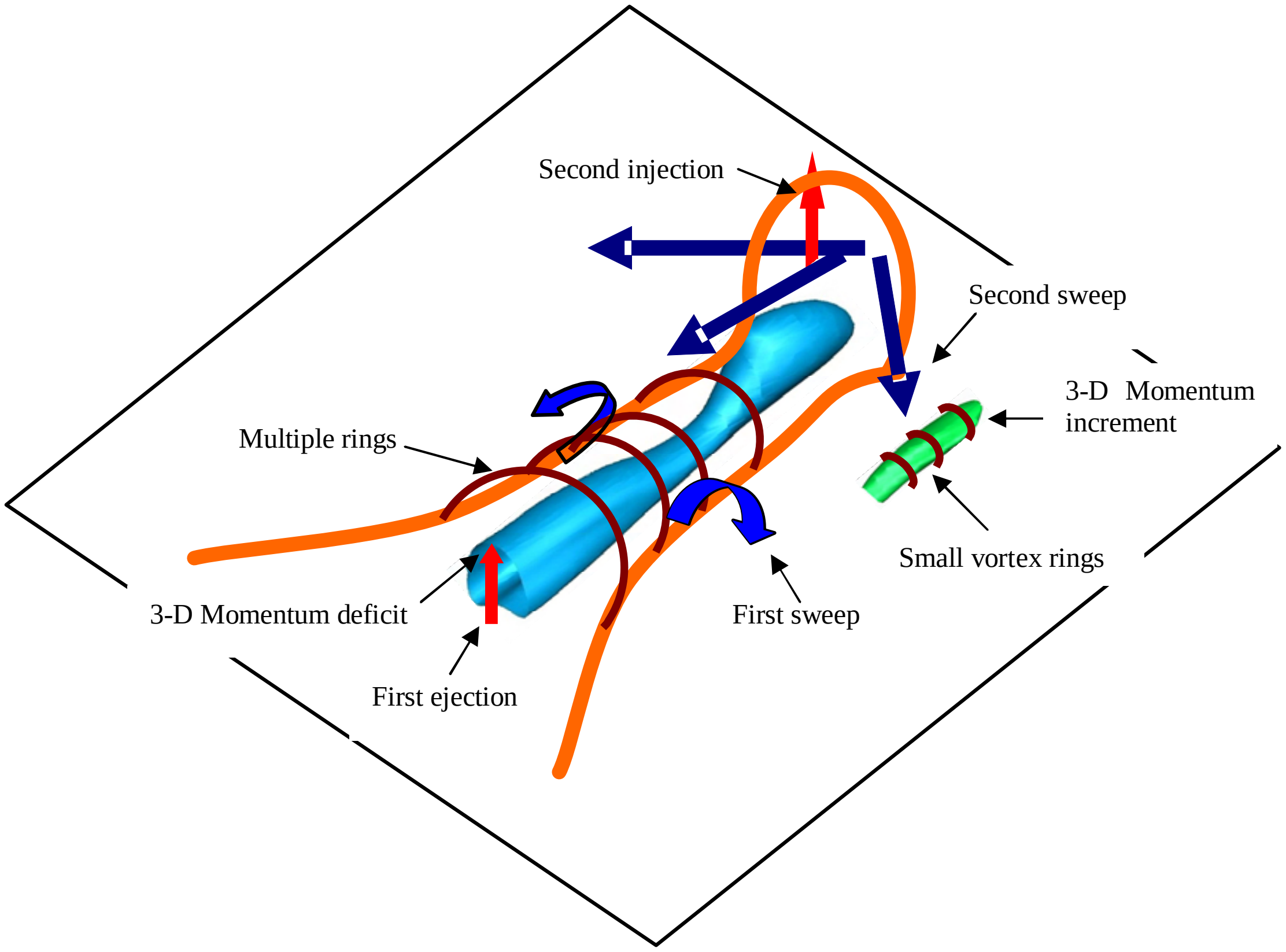

Figure 33: Sketch of mechanism of multiple rings formation and small vortices formation

*3.2.6* **Surface friction** *(new)*

Figure 34 is a time and spanwise averaged surface friction coefficient (CF). From the figure, it is clearly seen that there is a jump starting at x=430 which indicates an onset of flow transition and reach a maximum value at x=508.663. It is conventionally believed that the CF is large in the turbulent area and small in the laminar area due to the strong mixing in turbulent flow. That is the reason why most turbulence models are formulated based on change of turbulent viscosity. However, from Figure 34, it is easy to find that the CF reaches maximum in a laminar area and is not directly related to mixing. Further analysis found the viscosity is not changed for incompressible flow, but the shear layer is changed sharply when the first and second sweeps caused positive spikes (momentum increment), HS layers and consequent small vortices generations. The velocity gradients suddenly jump to a very large level in the laminar sub-layer and then the CF becomes very large. Therefore, the CF jump is not caused by mixing or viscosity coefficients increase, but is, pretty much, caused by velocity gradient jump due to small vortices formation.

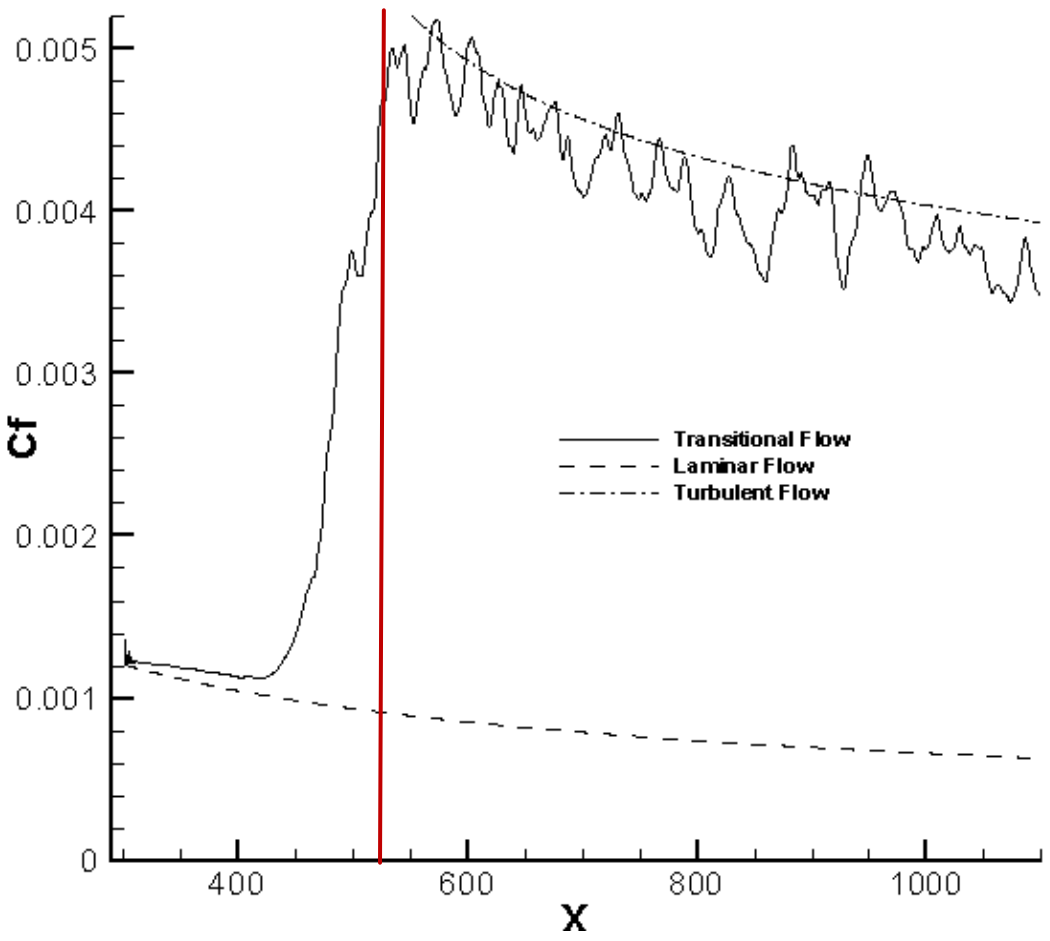

Figure 34: Maximum CF at x=508.633 (Laminar)

**3.3 Loss of symmetry (Randomization, new)**

The existing theory (Meyer et al, 2003) believes that the flow randomization in DNS is caused by the background noises and removal of periodic boundary conditions. They believe that randomization starts from loss of symmetry, which is caused by background noise. According to that theory, the ring tip is affected first, which rapidly influences the small length scale structure. This will lead to loss of symmetry and radomization for the whole flow field quickly.

However, what we observed in our new DNS is that the loss of symmetry starts from second level rings (Figure 35) while the top and bottom rings are still symmetric (Figure 36). The non-symmetric structure of second level rings will influence the small length scale at the bottom quickly. The change of symmetry in the bottom of the boundary layer is quickly spread to up level through ejections. This will lead to randomization of the whole flow field. Therefore, the internal instability of multiple level ring structure, especially the middle ring cycles, is a critical reason for flow randomization, but mainly not the background noise. In addition, the loss of symmetry starts in the middle of the flow domain, not the inflow and not the outflow (Figure 37a). As mentioned, we did not change the periodic boundary conditions and the solution is still periodic in the spanwise direction (period=2π). In addition, we did not add any additional background noise or inflow perturbation. However, we found that the flow lost symmetry first and then was randomized step by step (see Figure 38(a)-(d)). The flow was first periodic (period=π) and symmetric in the spanwise direction, $\sum_{k=0}^{n} a_k \cos(2ky)$, but then lost symmetry in some areas, and finally everywhere. It was periodic with a period of π, same as the inflow, but changed with a period of 2π. The flow is still periodic because we enforced the periodic boundary condition in the spanwise direction with a period of 2π. This means the flow does not only have $\sum_{k=0}^{n} a_k \cos(2ky)$ but also have $\sum_{k=0}^{n} b_k \sin(2ky)$ which is newly generated. Meanwhile, flow lost periodicity with period=π, but has to be periodic with period=2π (Figure 38(c) and 38(d)), which we enforced. Since the DNS study is focused on the mechanism of randomization and the DNS computation only allows use two periods in the spanwise direction, we consider that the flow is randomized when the symmetry is lost and period is changed from π to 2π (Figure 38(a) and 38(b)):

$$f(y) = \sum_{k=0}^{n} a_k \cos(2ky) + \sum_{k=1}^{n-1} b_k \sin(2ky) + \sum_{k=0}^{n-1} c_k \cos(2ky + y) + \sum_{k=0}^{n-1} d_k \sin(2ky + y)$$

In real flow, there is no such a restriction of periodic boundary condition in the spanwise direction. A more detailed discussion about the mechanism of flow randomization will be given by another AIAA 2012 paper (Lu et all, 2012a). C. Liu gave a hyphothesis that the loss of symmetry may be caused by C- and K-type transition shift, but it has to be checked by further investigation.

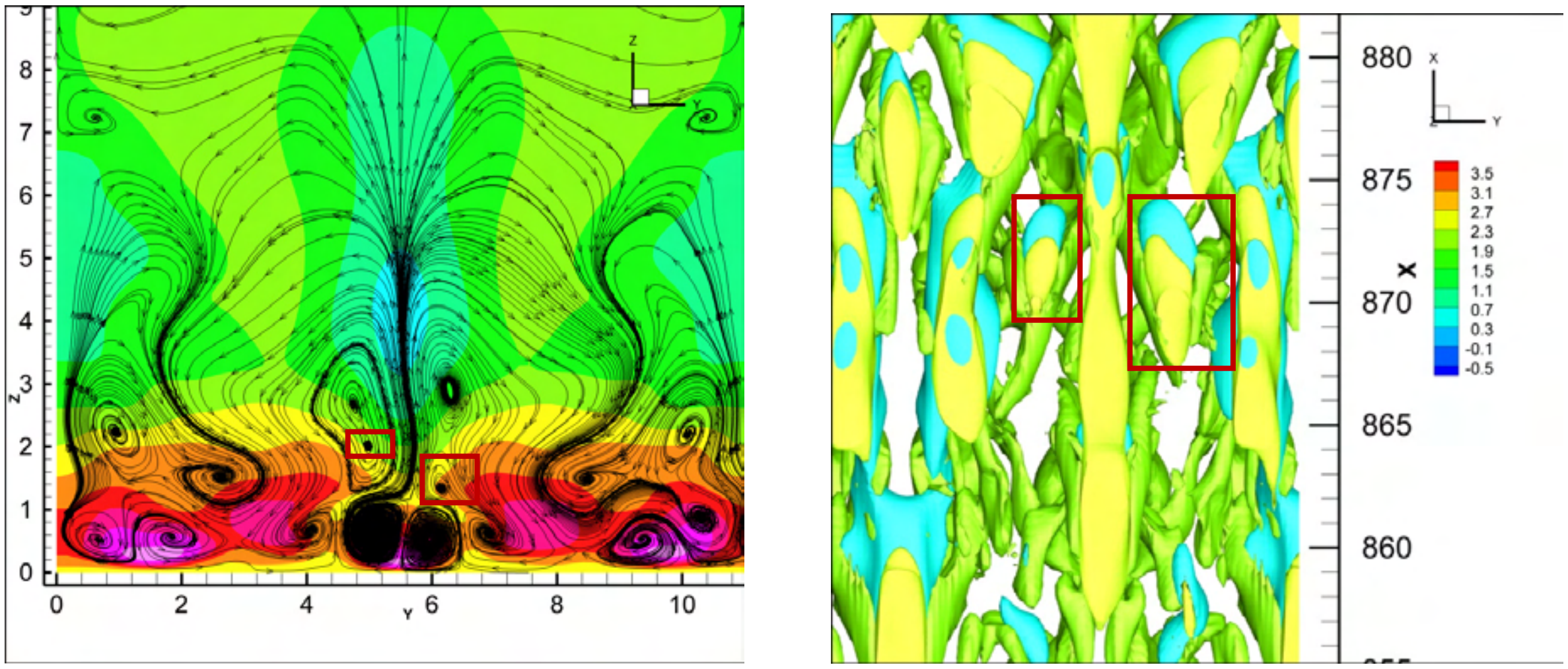

(a) Section view in y-z plane (b) Bottom view of positive spike

Figure 35: The flow lost symmetry in second level rings and bottom structure at t=15.0T

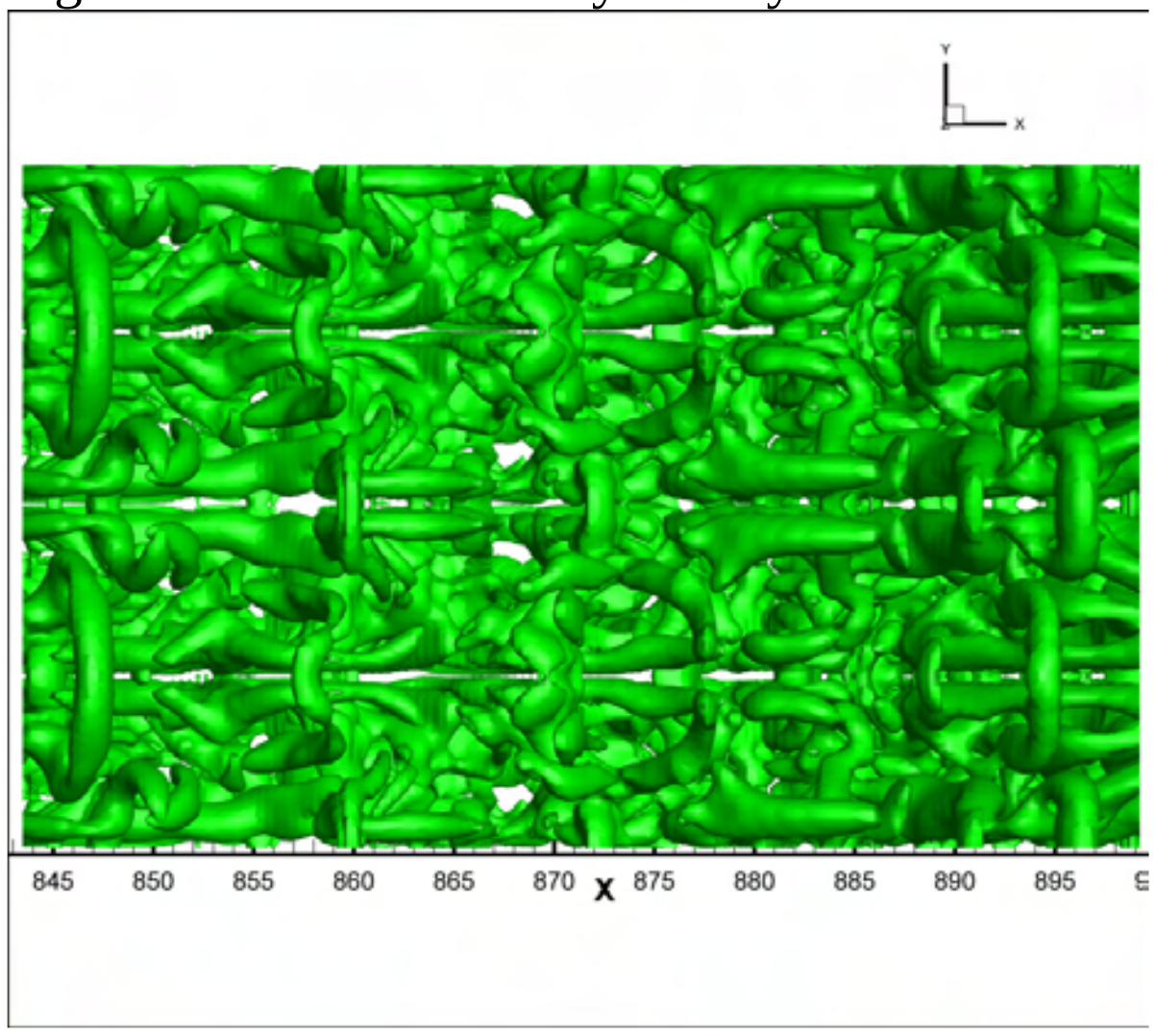

Figure 36: The top ring structure is still symmetric at t=15.0T

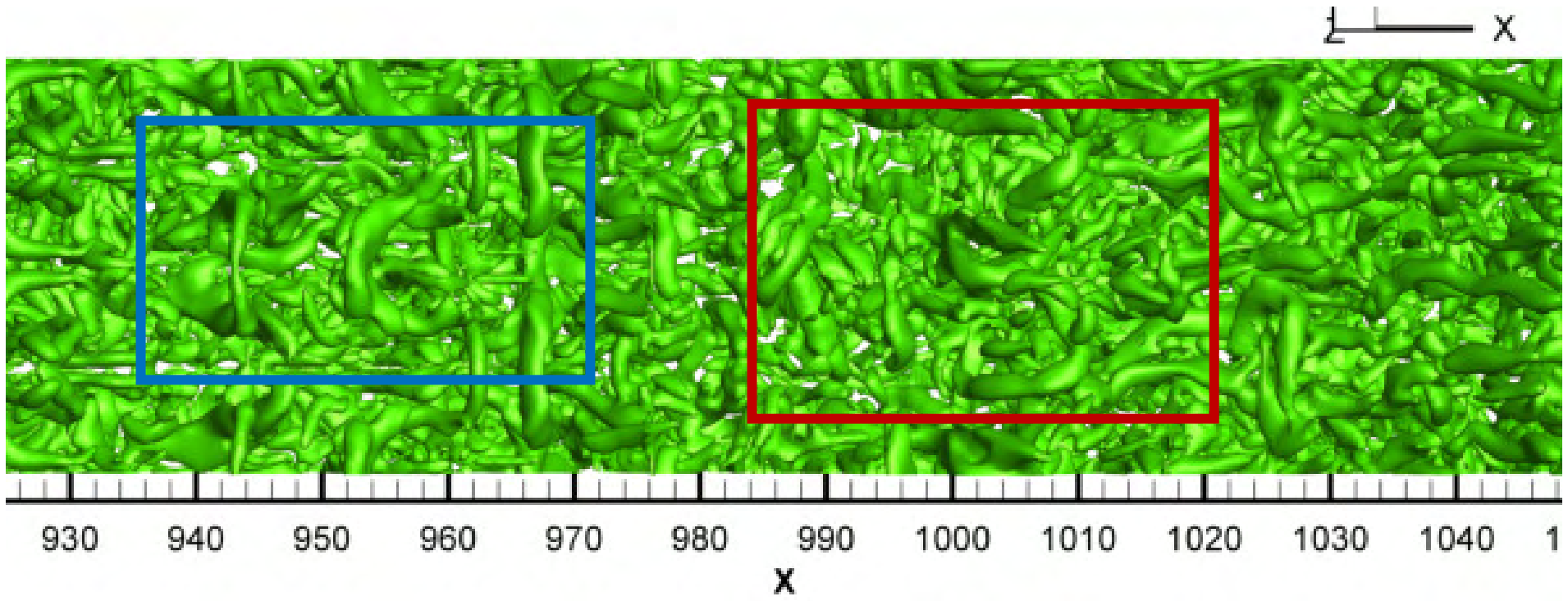

(a) Top ring structure lost symmetry (blue area is symmetric but red area is not)

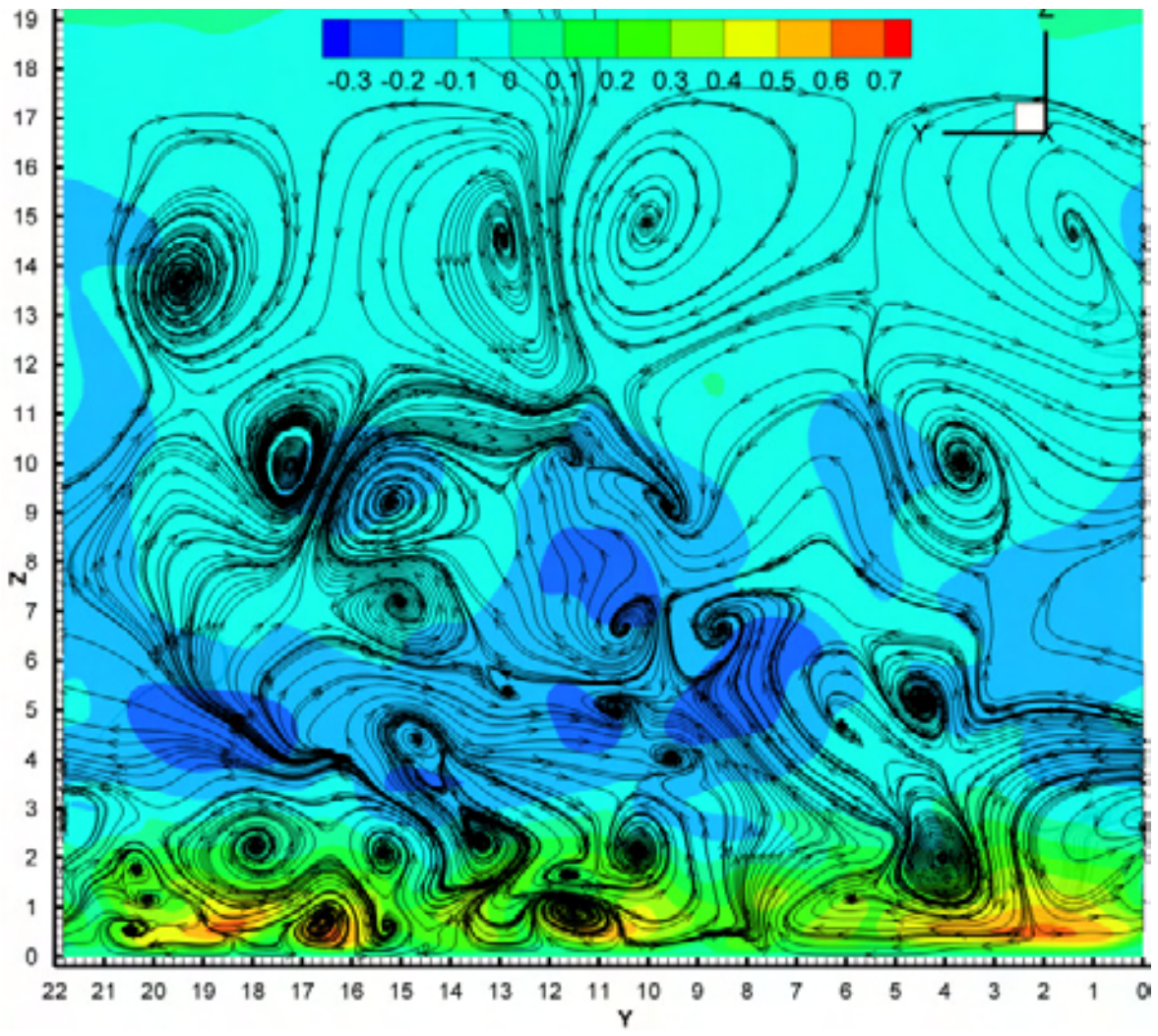

(b) Symmetry loss in the whole section of y-z plane

Figure 37: The whole flow field lost symmetry at t=21.25T

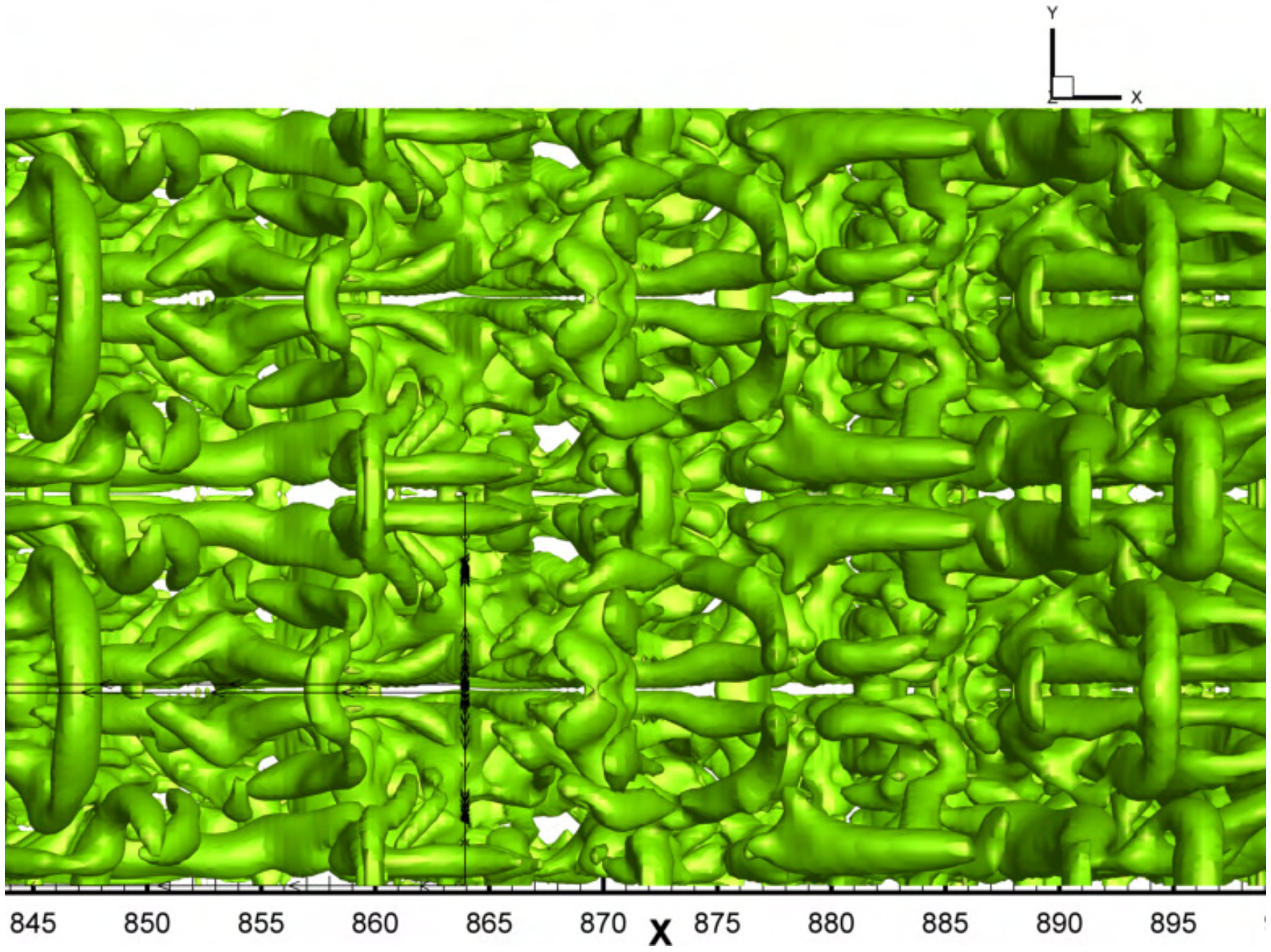

Figure 38a: Whole domain is symmetric and periodic – a × cos 2y (the period=π; spanwise domain is –π<y<π)

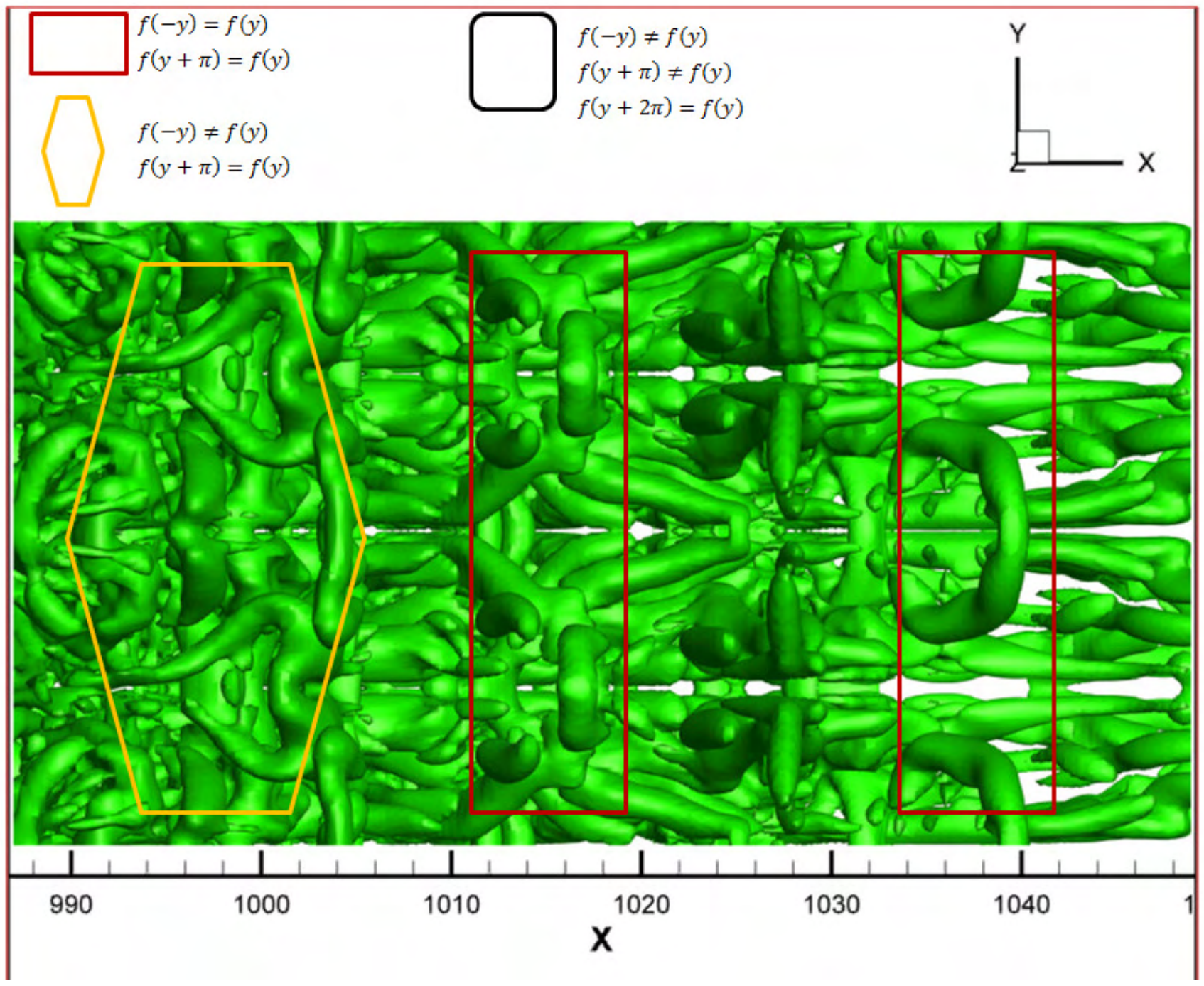

Figure 38b: Symmetric and asymmetric – red rectangular: periodic and symmetric at y=-π/2, 0, π/2, i.e. f(-π/2 –y)=f(-π/2+y), f(-y)=f(y), f(π/2-y)=f(π/2+y); yellow diamond: periodic, f(y+π)=f(y), period=π; but asymmetric $f(-y) \neq f(y)$ ; the spanwise domain is –π<y<π

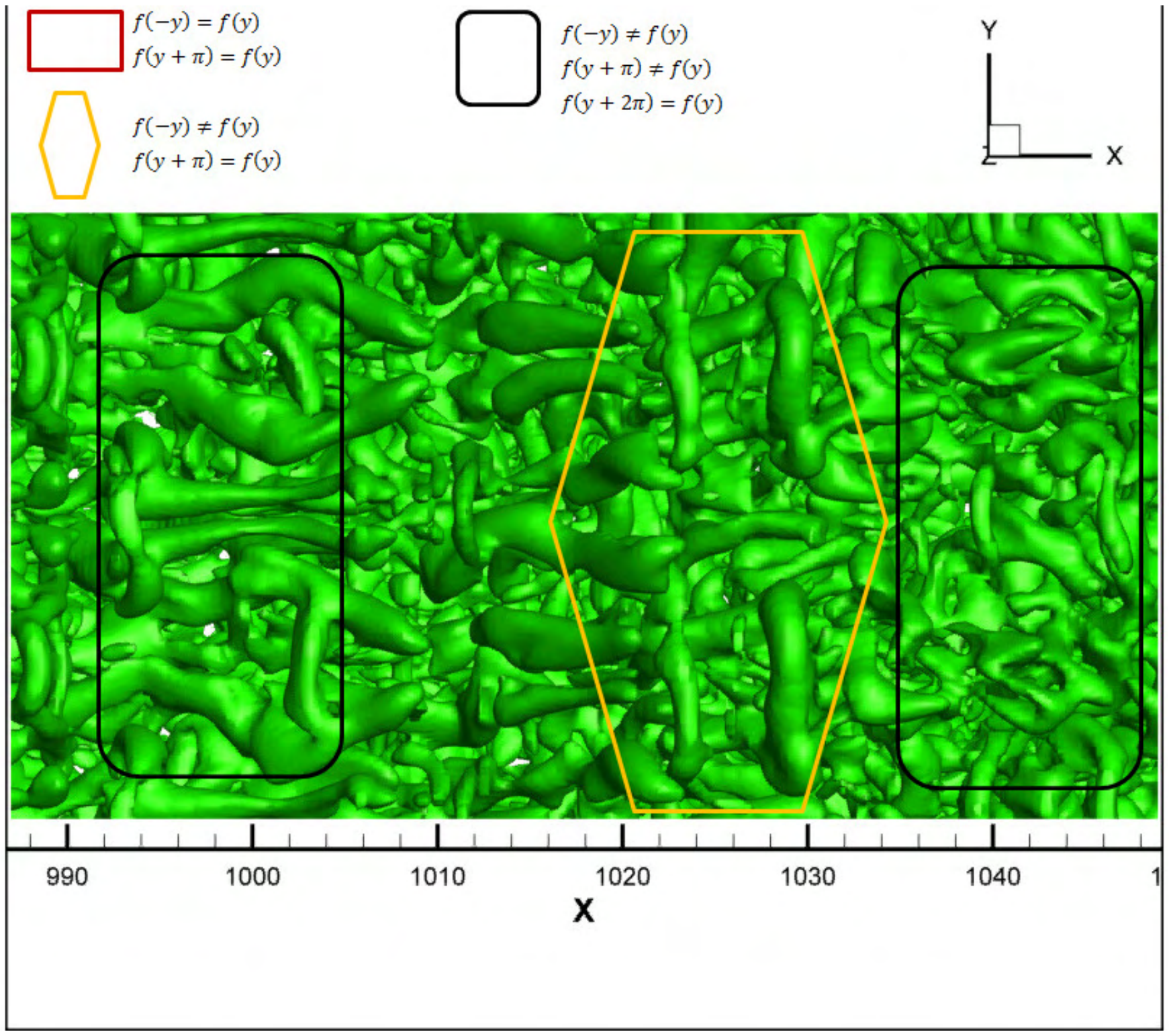

Figure 38c: Periodic but asymmetric – yellow diamond: periodic, period=π; black box: periodic but period= 2π; the spanwise domain is –π<y<π

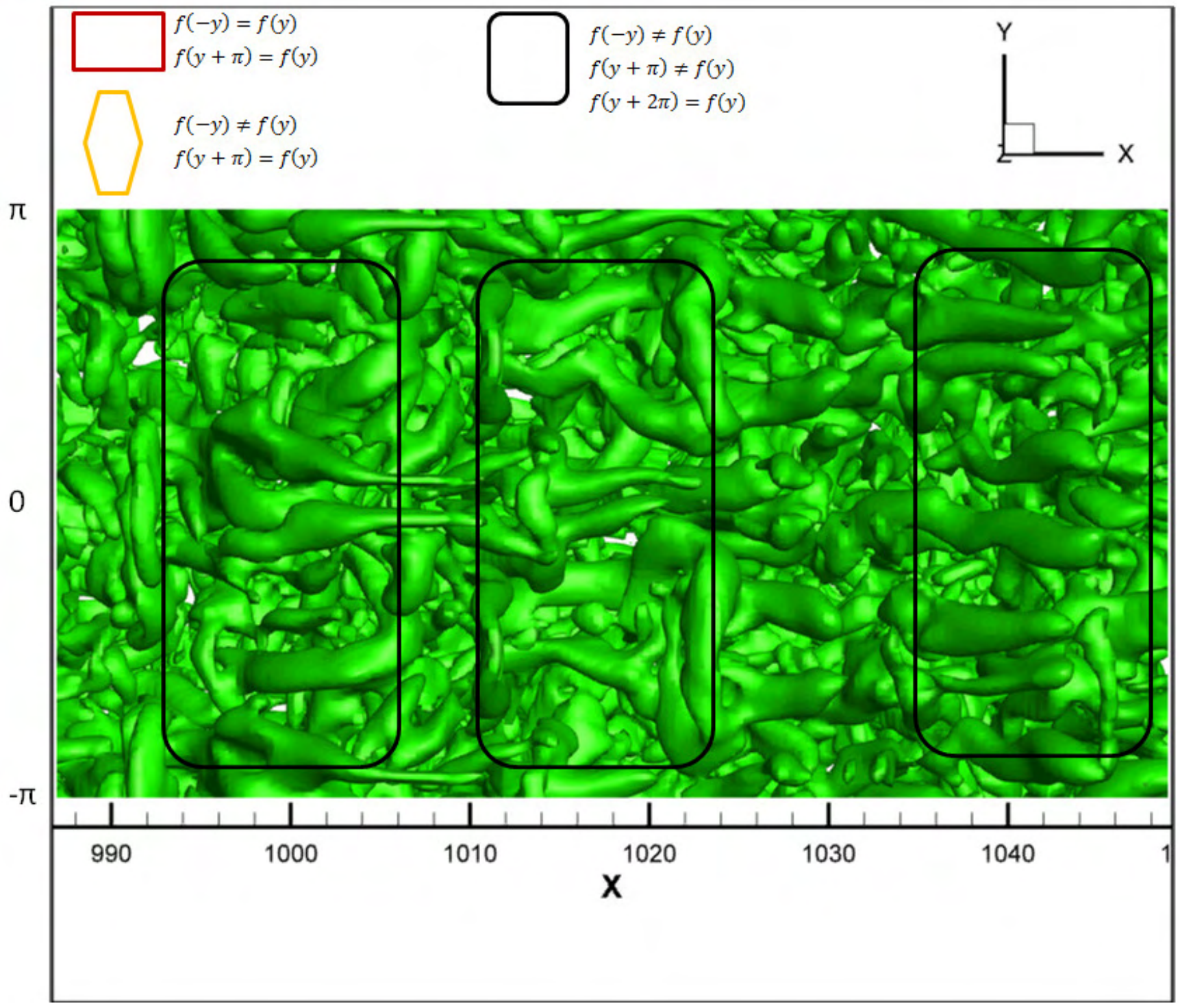

Figure 38d: Periodic but asymmetric – all black boxes: periodic but asymmetric (period= 2π); the spanwise domain is –π<y<π

## IV. Summary of Our DNS Findings

### 4.1 Our new observations

Since turbulence generation and sustenance are one of the top secret in nature, this research will bring significant impact on fundamental fluid mechanics as the classical theory and dominant concepts on late flow transition and turbulence structure are challenged. Table 2 gives a comparison of classical theory and currently dominant conclusions with the observation by our DNS:

Table 2 comparison of classical theory and our DNS observation

| **Topic** | **Classical or Existing Theory** | **Observation of Our DNS** |
|---|---|---|
| Turbulence generation | By “Vortex Breakdown” | Not by “Vortex Breakdown” but by shear layer instability |
| First ring generation | Self-induced, deformed, inclined and pinched-off | By counter-rotated vortices interaction, circular, perpendicular, no pinch-off |
| Multiple ring structure | “Crow theory” or breakdown and then re-connected | No breakdown, not “Crow theory” but momentum deficit caused by ejection, vorticity conservation |
| Multiple level high shears | No report | By multiple level sweeps and ejections |
| Energy transfer channel, turbulence sustenance | energy transfers from larger vortices to smaller one through “vortex breakdown” without dissipation until viscosity | From inviscid flow down to bottom by multilelvel sweeps |
| U-shaped vortex | Head wave, secondary vortex, by second sweep, newly formed, breakdown | Not head wave, tertiary vortex, by secondary vortex, existing from beginning, never breakdown |
| Randomization | Background noise, starting from the top ring and then going down to the bottom | Internal property, starting from second level rings in the middle, affects bottom and then up to affect top rings. Loss of symmetry maybe caused by C-K shift. |
| Coefficients of friction | Turbulent flow has large friction due to strong boundary layer mixing | Depending only on velocity profile changes in laminar sub-layer, no matter turbulent or laminar |
| Richardson eddy cascade | Classical theory | Not observed |
| Vortex breakdown | Classical theory | Not observed |
| Kolmogorov scale | Classical theory | Smallest length scale should be determined by minimum shear layer instability |

## 4.2 New theory on turbulence formation and sustenance by C. Liu

4.2.1 Classcal theory on turbuluce

A turbulent flow is characterized by a hierarchy of scales through which the energy cascade takes place. According to Richardson, turbulence is random interactions of "eddies" as "**big whorls have little whorls, little whorls have smaller whorls, that feed on their velocity, and so on to viscosity**" (Figure 40 by Feynman, 1955). On the other hand, vortex stretching is the core mechanism on which the turbulence energy cascade relies to establish the structure function. According to Richardson (1928) and Kolmogorov (1941), the radial length scale of the vortices decreases and the larger flow structures break down into smaller structures (Figure 41 by Frisch et al, 1978). The process continues until the small scale structures are small enough to the extent where their kinetic energy is overwhelmed by the fluid's molecular viscosity and dissipated into heat.

Richardson’s energy cascade theory has several important points of view which can be summarized as follows:

1. Stremwise vortex is stretching
2. Vortex must break down to smaller structure due to the stretching
3. Energy transfers from large vortex to smaller one through “vortex breakdown”
4. These smaller vortices continue the same process until the smallest one appears (Kolmgorov’s length scale) in which viscosity is dominant
5. The micro-scale η (Kolmogorov scale) is several order smaller than the macro-scale L (main flow scale) and the middle length scales r (η<<r<<L) are basically inertial.

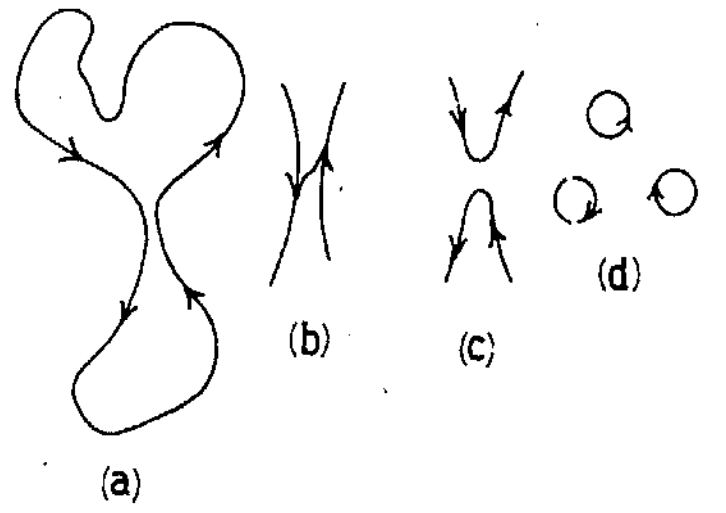

Fig. 10. A vortex ring (a) can break up into smaller rings if the transition between states (b) and (c) is allowed when the separation of vortex lines becomes of atomic dimensions. The eventual small rings (d) may be identical to rotons.

Figure 40: Sketch of vortex breakdown (Feynman, 1955; Tsubota et al, 2009)

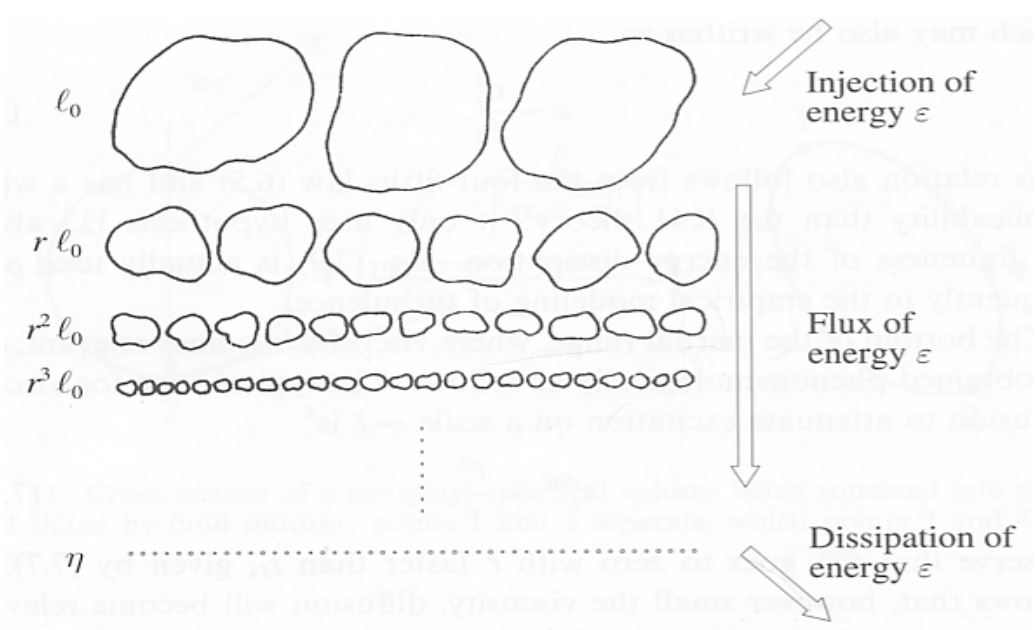

Figure 41: Sketch of Richardson cascade process (Frisch et al, 1978)

### 4.2.2 New theory on turbulence formation and sustenance by C. Liu

On the other hand, our new theory can be described as follows:

1. Stremwise vortex is stretched but never breaks down.
2. Vortex has strong rotation which causes strong mixing
3. In the middle of two vortex legs, there is a low speed zone formed by ejection from boundary layer bottom to bring momentum deficit up to this area. This momentum deficit forms a long shear layer
4. This shear layer in the middle of two legs is unstable and multiple rings with spanwise vorticity will form one by one to generate multiple rings near the inviscid region due to the shear layer instability. In this process, Helmholtz vorticity conservation must be satisfied.
5. In the two sides of the vortex legs, high speed zones are formed by sweeps which bring high energy (energy increment) from inviscid area to the near wall region. The energy increment will form shear layers near the wall
6. The momentum increment shear layers near the wall are unstable (shear layer instability) and will form multiple smaller ring-like vortices
7. These newly generated large and smaller vortices will continue similar process to generate smaller ring-like vortices until the vortex size is too small and the new shear layer become stable. The shear layer stability analysis should be further studied. This will be reported by other papers of ours.
8. Energy is transferred from the large size vortices to small size vortices through multiple level sweeps. Dissipation is inevitable during the energy transfer.
9. All smaller vortices are newly generated by shear layer instability, but not by original vortex "breakdown"
10. Smallest length scales are all generated by wall surface which must related to the surface configuration and cannot be "isotropic."
11. Surface drag is determined by small length scale generation, which must be closely related to turbulence modeling
12. "Shear layer" control is the key issue for flow transition control, turbulent flow control, and drag reduction

Of course, the new theory has to be further studied and confirmed.

## 4.3 Possible impact on fluid mechanics

### 4.3.1 Impact on fundamental fluid mechanics

Since turbulence generation and sustenance are one of the top mysteries in nature, this research will bring significant impact on fundamental fluid mechanics as the classical theories on late flow transition and turbulence structure are challenged. LBLT classical theory, Richardson eddy cascade, Kolmogorov first and second hypotheses should be all revisited.

### 4.3.2 Impact on DNS and LES

As shown by the new DNS, it is not observed that "Kolmogorov scale" is the smallest turbulence scale but is found that the smallest length scale is determined by "shear layer instability". The estimate of energy containing scale, $\frac{_{\min}}{_{ref}} = (\frac{1}{\mathrm{Re}})^{3/4}$, is too conservative and has no solid foundation. For example, an engineering hydrodynamics problem has commonly a Reynolds number, $\mathrm{Re} = 10^8$. Considering 3-D in space and 1-D in time, the requirement for computation would be $Q = \mathrm{Re}^{4\times 3/4} = (10^8)^3 = 10^{24}$. This is apparently impossible for any currently available computer or any near future computers. However, researchers in US National Research Labs and many others around the world do "implicit LES" which is really “coarse grid DNS”. They do DNS, but are afraid to call DNS but “ILES”. The reason is that they are afraid with “Kolmogorov scale”. However, our DNS already find that “Kolmogorov scale” does not have solid scientific foundation and must be re-visited. This will provide guidance to DNS which is supposed to resolve all energy-contained small length scales and LES which is expected to resolve the large eddies and model the subgird scales. Our new findings may make DNS/LES useable to some engineering applications.

#### 4.3.3 Impact on turbulence modeling

The eddy viscosity turbulence model is mainly based on “Boussinesq Assumption” (1972) that the non-linear perturbation terms can be modeled by resolved mean scales with new turbulent coefficients of viscosity. However, most turbulence models are still empirical and case-related. The right turbulence models can be developed with deep understanding of turbulence generation and sustenance. As shown by our DNS, the perturbation and small length scales are not isotropic but configuration related. This will give new concepts for “turbulence modeling.”

#### 4.3.4 Impact on flow control of flow transition and turbulence

Flow transition and turbulence control should be based on right understanding on flow transition and turbulence structure. Based on our observation, it is proposed to cut the energy transfer channels for high shear layer generation to stop or delay turbulence generation. It is also found that the “shear layer instability” is the “mother of turbulence.” Apparently, control of shear layer formation and structure is the key issue to the control of the transitional flow and fully developed turbulent flow. It is also an important way for drag production of turbulent flow, e.g. why oscillating wall and rough wall with riblets can reduce the surface drag.

## V. Conclusions

In our DNS work (Chen et al., 2009, 2010a, 2010b, 2011a, 2011b; Liu et al., 2010a, 2010b, 2010c, 2011a, 2011b 2011c, 2011d; Lu et al., 2011a, 2011b, 2011c), we have made a number of new findings on late flow transition in a boundary layer including the mechanism of large vortex structure formation, small-length scale generation and flow randomization (loss of symmetry), which can be summarized as follows (some of conclusions have been published in our previous papers):

1. Large vortex structure formation:

1) The widely spread concept, “hairpin vortex breakdown to small pieces”, which was considered as the last step of flow transition, is not observed and is found incorrect (Liu et al, 2011c, 2011d)

2) The ring-like vortex with or without legs is found to be the only form existing inside the flow field (Liu et al, 2011c)

3) First vortex ring formation is the result of the interaction between two pairs of counter-rotating primary and secondary streamwise vortices (Liu et al, 2011b)

4) The formation of the multiple ring structure is caused by momentum deficit and consequent shear layer. The deficit is induced by primary vortex leg ejection which brings the low speed flow from the boundary layer bottom up.

5) Multiple ring generation also follows the first Helmholtz vortex conservation law. When the ring is generated, the primary streamwise vorticity is reduced.

6) The U-shaped vortex is not a secondary vortex induced by second sweeps and is not newly formed as the head of young turbulence spot and never breaks down to small pieces. Actually, the U-shaped vortex is induced by the secondary vortices but not by second sweeps. The U-shaped vortex is a vortex tube but not heading wave. In addition, it is found that the U-shaped vortex has the same vorticity sign as the original λ-shaped vortex tube legs, which means the U-shaped vortex is not secondary but tertiary, and serves as an additional channel to support the multiple ring-like vortices (Lu et al, 2011a).

7) Neither the hairpin vortex nor U-shaped vortex can break down. The existing articles, which reported vortex breakdown, were either based on 2-D visualization or using low pressure center as the vortex core (Liu et al, 2011d)

8) The U-shaped vortex becomes distinguishable when the heading ring is skewed and sloped. Consequently, the second sweep becomes weak and positive spike disappears. In such a case, the small length scales are damped quickly by the strong boundary layer viscosity due to the lack of energy supplies by second sweeps from the inviscid region (Liu et al 2011d)

9) The leading rings are almost perfectly circular which are not deformed because they are located near the inviscid region which is isotropic. The leading rings stand almost perpendicularly not with 45 degrees because both the top and the bottom of the ring are located in the upper bound of the boundary layer and move at a same speed (U=1). Some later rings do not stand perpendicularly because they are not located fully in the inviscid region. The top of the ring moves faster than the bottom of the ring due to the boundary layer mean velocity profile. The ring legs are inclined because they are located inside the boundary layer and must be stretched and inclined (Liu et al 2011b, 2011d).

10) The momentum increment (positive spikes) moves very fast at the same speed as the negative spikes (original vortex rings). This is because the positive spike is produced by first and second sweeps which bring high speed flow from the inviscid region. There is "no shock waves" as some research papers suggested.

11) The momentum increment (positive spikes) look like Olive shape.

2. The small length scale generation

1) The small vortices can be found on the bottom of the boundary layer near the wall surface (bottom of the boundary layer). It is justified that all small vortices, and thus turbulence, are generated by high shear layers near the wall surface (Liu et al 2011d; Lu et al 2011b).

2) The negative (momentum deficit) and positive spikes (momentum increment) generate high shear (HS) layers near the wall due to zero speed on the wall and large speed in the inviscid zones.

3) HS is unstable especially for those near the wall surface and HS can form many small vortices due to HS instability. Each of small vortex rings has two legs, which lie down on the wall surface.

4) The new vortex rings can generate new second sweeps and form newer negative and positive spikes, which is called by us as “multiple level second sweeps” and “multiple level negative spike (momentum deficit) and positive spikes (momentum increment)”, which are located even lower and closer to the wall surface.

5) Energy can be transferred from the inviscid area to the bottom of the boundary layer through the mechanism with “multiple level sweeps and multiple level positive spikes” (Liu et al 2011d; Lu et al 2011b)

6) All small vortices are generated around the high shear layers. The mechanism of formation of multiple level negative and positive spikes is one of the most important concepts in understanding the physical nature of energy transformation.

7) When the leading ring is deformed and/or the standing position is inclined, the second sweep and then the positive spikes will be weakened. Then the small length scales could quickly damp and the large coherent vortex structure becomes distinguishable (turbulence spot head) (Lu et al 2011a)

8) The mechanism of small length scale (turbulence) generation can be described by following chart: perfect circular and perpendicular vortex rings generate strong second sweeps, and then second sweeps bring the high energy from the inviscid region to the bottom of the boundary layer and generate the strong positive spikes. The positive spikes generate high shear layers which are not stable, and so that HS and wall surface generate more and more small length scales near the wall surface (Lu et al 2011b).

9) The surface friction suddenly jumped when the small length scale forms due to the positive spikes. Therefore the surface friction is related to small length scale generation which changes the velocity profile in the boundary sublayer but not directly related to mixing

3. Loss of symmetry (Randomization)

The randomization starts from loss of symmetry in the middle of multiple circles of vortex ring structure, not on the top rings and not on the bottom. The randomization is an internal property not only caused by big background noises or removal of periodic boundary conditions in DNS.

4. Comments on classical theory of turbulence

1) There is no DNS evidence supporting Richardson's vortex cascade hypothesis and the "vortex breakdown" cannot happen

2) There is no DNS evidence supporting Kolmogorov's assumption on "isotropic small vortices" and "non-dissipative energy passing from larger vortices to smaller vortices"

3) There is no DNS evidence supporting "Kolmogorov's small length scale.

5. Comments on turbulence modeling

Although Prandtl's "mixing length" model is an old concept (Prandtl 1925), but turbulence is really generated by mixing of variety of different size vortices and the "mixing length" concept coincides with physics. The serious question is to find the right size of "mixing length". We will develop a new "mixing length" model based on DNS data and "shear layer stability" analysis.

## Acknowledgments

This work was supported by AFOSR grant FA9550-08-1-0201 supervised by Dr. John Schmisseur and the Department of Mathematics at University of Texas at Arlington. The authors are grateful to Texas Advanced Computing Center (TACC) for providing computation hours. This work is accomplished by using Code DNSUTA which was released by Dr. Chaoqun Liu at University of Texas at Arlington in 2009.

## References

[1] Adrian, R. J., Hairpin vortex organization in wall turbulence, Physics of Fluids, Vol 19, 041301, 2007

[2] Bake S, Meyer D, Rist U. Turbulence mechanism in Klebanoff transition: a quantitative comparison of experiment and direct numerial simulation. J.Fluid Mech, 2002 , 459:217-243

[3] Boroduln V I, Gaponenko V R, Kachanov Y S, et al. Late-stage transition boundary-Layer structure: direct numerical simulation and experiment. Theoret.Comput.Fluid Dynamics, 2002,15:317-337.

[4] Boussinesq, J. Essai sur la theorie des eaux courantes. Memories Aca,. Des Science, Vol. 23, No. 1 Paris, 1872

[5] Chen, L., Liu, X., Oliveira, M., Tang, D., Liu, C., Vortical Structure, Sweep and Ejection Events in Transitional Boundary Layer, Science China, Series G, Physics, Mechanics, Astronomy, Vol. 39 (10) pp1520-1526, 2009

[6] Chen, L., Liu, X., Oliveira, M., Liu, C., DNS for ring-like vortices formation and roles in positive spikes formation, AIAA Paper 2010-1471, Orlando, FL, January 2010a.

[7] Chen L., Tang, D., Lu, P., Liu, C., Evolution of the vortex structures and turbulent spots at the late-stage of transitional boundary layers, Science China, Physics, Mechanics and Astronomy, Vol. 53 No.1: 1–14, January 2010b,

[8] Chen, L., Liu, C., Numerical Study on Mechanisms of Second Sweep and Positive Spikes in Transitional Flow on a Flat Plate, Journal of Computers and Fluids, Vol 40, pp28-41, 2011a

[9] Chen, L., Liu, X., Tang, D., Liu, C. Evolution of the vortex structures and turbulent spots at the late-stage of transitional boundary layers. Science of China, Physics, Mechanics & Astronomy, 2011 Vol. 54 (5): 986-990, 2011b

[10] Crow S C. Stability theory for a pair of trailing vortices. AIAA J, 1970, 8: 2172-2127

[11] Davidson, P. A., *Turbulence: An Introduction for Scientists and Engineers*. Oxford University Press.

ISBN 9780198529491, 2004
[12] Duros F, Comte P, Lesieur M. Large-eddy simulation of transition to turbulence in a boundary layer developing spatially over a flat plate. J.Fluid Mech, 1996, 326:1-36
[13] Frisch, U., *Turbulence: The Legacy of A. N. Kolmogorov*. Cambridge University Press, 1995
[14] Frisch, U., Sulem, P.L., and Nelkin, M., A simple dynamical model of intermittent fully developed turbulence, Journal of Fluid Mechanics , Volume 87, Issue 04, pp 719 – 736, 1978
[15] Feynman, R. F., in *Progress in Low Temperature Physics*, vol. **1**, Chap.2 C. J. Gorter (ed), North Holland Publishing Co., Amsterdam (1955).
[16] Guo, Ha; Borodulin, V.I..; Kachanov, Y.s.; Pan, C; Wang, J.J.; Lian, X.Q.; Wang, S.F., Nature of sweep and ejection events in transitional and turbulent boundary layers, J of Turbulence, August, 2010
[17] Herbert, T., 1988, "Secondary Instability of Boundary Layer," Annu. Rev.Fluid Mech., 20, pp. 487-526.
[18] Jeong J., Hussain F. On the identification of a vortex, J. Fluid Mech. 1995, 285:69-94
[19] Jiang, L., Chang, C. L. (NASA), Choudhari, M. (NASA), Liu, C., Cross-Validation of DNS and PSE Results for Instability-Wave Propagation, AIAA Paper #2003-3555, The 16th AIAA Computational Fluid Dynamics Conference, Orlando, Florida, June 23-26, 2003
[20] Kachnaov, Y. S., 1994, "Physical Mechanisms of Laminar-Boundary-Layer Transition," Annu. Rev. Fluid Mech., 26, pp. 411–482.
[21] Kachanov, Y.S. On a universal mechanism of turbulence production in wall shear flows. In: Notes on Numerical Fluid Mechanics and Multidisciplinary Design. Vol. 86. Recent Results in Laminar-Turbulent Transition. — Berlin: Springer, 2003, pp. 1–12.
[22] Kleiser L, Zang T A. Numerical simulation of transition in wall-bounded shear flows. Annu.Rev.Fluid Mech.1991.23:495-537
[23] Kloker, M and Rist U. , Direct Numerical Simulation of Laminar-Turbulent Transition in a Flat-Plate Boundary Layer on line at http://www.iag.uni-stuttgart.de/people/ulrich.rist/publications.html
[24] Kolmogorov, Andrey Nikolaevich (1941). "The local structure of turbulence in incompressible viscous fluid for very large Reynolds numbers". *Proceedings of the USSR Academy of Sciences* **30**: 299–303. **(Russian)**, translated into English by Kolmogorov, Andrey Nikolaevich (July 8, 1991). "The local structure of turbulence in incompressible viscous fluid for very large Reynolds numbers". *Proceedings of the Royal Society of London, Series A: Mathematical and Physical Sciences* **434** (1991): 9–13. Bibcode 1991RSPSA.434....9K. doi:10.1098/rspa.1991.0075.
[25] Lee C B., Li R Q. A dominant structure in turbulent production of boundary layer transition. Journal of Turbulence, 2007, Volume 8, N 55
[26] Liu, X., Chen, L., Oliveira, M., Tang, D., Liu, C., DNS for late stage structure of flow transition on a flat-plate boundary layer, AIAA Paper 2010-1470, Orlando, FL, January 2010a.
[27] Liu, C., Chen, L., Study of mechanism of ring-like vortex formation in late flow transition, AIAA Paper 2010-1456, Orlando, FL, January 2010b.
[28] Liu, X., Chen, Z., Liu, C., Late-Stage Vortical Structures and Eddy Motions in Transitional Boundary Layer Status, Chinese Physics Letters Vol. 27, No.2 2010c
[29] Liu, C., Chen, L., Lu, P., New Findings by High Order DNS for Late Flow Transition in a Boundary Layer, J of Modeling and Simulation in Engineering, to appear, open access journal and copy right is kept by authors, 2011a
[30] Liu, C., Chen, L., Parallel DNS for vortex structure of late stages of flow transition, J. of Computers and Fluids, Vol.45, pp 129–137, 2011b
[31] Liu, C., Numerical and Theoretical Study on "Vortex Breakdown", International Journal of Computer Mathematics, to appear, 2010c, on line http://www.tandfonline.com/doi/abs/10.1080/00207160.2011.617438
[32] Liu, C., Chen, L., Lu, P., and Liu, X., Study on Multiple Ring-Like Vortex Formation and Small Vortex Generation in Late Flow Transition on a Flat Plate, Theoretical and Numerical Fluid Dynamics, to appear, 2010d, on line http://www.springerlink.com/content/e4w2q465840nt478/

[33] Lu, P., Liu, C., Numerical Study of Mechanism of U-Shaped Vortex Formation, AIAA Paper 2011-0286, and Journal of Computers and Fluids, to appear January 2011a , on line at http://authors.elsevier.com/TrackPaper.html?trk_article=CAF1706&trk_surname=Liu
[34] Lu, P. and Liu, C., DNS Study on Mechanism of Small Length Scale Generation in Late Boundary Layer Transition, AIAA Paper 2011-0287 and J. of Physica D, Non-linear, 241 (2012) 11-24, 2011b, on line: http://www.sciencedirect.com/science/article/pii/S0167278911002612
[35] Lu, P., Thampa, M, Liu, C., umerical Study on Randomization in Late Boundary Layer Transition, AIAA 2012 Scientific Meeting, Nashville, TN, January, 2012a
[36] Marshak, Alex, *3D radiative transfer in cloudy atmospheres; pg.76*. Springer. ISBN 9783540239581. http://books.google.com/books?id=wzg6wnpHyCUC, 2005
[37] MEYER, D.G.W.; RIST, U.; KLOKER, M.J. (2003): Investigation of the flow randomization process in a transitional boundary layer. In: Krause, E.; Jäger, W. (eds.): *High Performance Computing in Science and Engineering '03.* Transactions of the HLRS 2003, pp. 239-253 (partially coloured), Springer.
[38] Moin, P., Leonard, A. and Kim, J., Evolution of curved vortex filament into a vortex ring. *Phys. Fluids*, **29**(4), 955-963, 1986
[39] Mullin, Tom, Turbulent times for fluids, *New Scientist*., 11 November 1989
[40] Prandtl, L., Bericht uber Untersuchungen zur ausgebildeten Turbulenz. ZAMM. Z. angew. Math. Mech., Bd. 5. 136-139, 1925
[41] Rist, U. and Kachanov, Y.S., 1995, Numerical and experimental investigation of the K-regime of boundary-layer transition. In: R. Kobayashi (Ed.) *Laminar-Turbulent Transition* (Berlin: Springer) pp. 405-412.
[42] Sandham, D. and Kleiser, L., The late stages of transition in channel flow, J. Fluid Mech., Vol 245, pp. 319-348, 1992[24] Schlichting, H. and Gersten, K., Boundary Layer Theory, Springer, 8th revised edition, 2000
[43] Singer, B. and Joslin, R., Metamorphosis of a hairpin vortex into a young turbulent spot, Phys. Fluids, 6 (11), November 1994, pp 3724-3736, 1994.
[44] *USA Today*, Turbulence theory gets a bit choppy, September 10, 2006. http://www.usatoday.com/tech/science/columnist/vergano/2006-09-10-turbulence_x.htm.
[45] Wu, X. and Moin, P., Direct numerical simulation of turbulence in a nominally zero-pressure gradient flat-plate boundary layer, JFM, Vol 630, pp5-41, 2009